\documentclass[reprint, superscriptaddress, groupedaddress, a4paper, nobibnotes,floatfix, amsmath, amssymb, twocolumn, pre, showkeys,balancelastpage]{revtex4-2}

\usepackage[utf8]{inputenc}
\usepackage[T1]{fontenc}
\usepackage[english]{babel}
\usepackage{cancel}
\usepackage{amssymb}
\usepackage{bigints}
\usepackage{color}
\usepackage{xcolor}
\usepackage{hyperref}
\usepackage{comment,soul}

\usepackage{booktabs}
\usepackage{amsmath,amsfonts,bm,mathrsfs,dsfont,amssymb,xfrac,bigints,cancel,physics,mathtools}
\usepackage{braket}

\hypersetup{colorlinks=true, linkcolor=blue, linktoc=page, citecolor=blue, urlcolor=blue}

\renewcommand{\vec}[1]{\mathbf{#1}}
\newcommand{\mat}[1]{\bm{\mathsf{#1}}}

\begin{document}
\title{Nonlinear density waves on graphene electron fluids}
\author{Pedro Cosme}
\email{pedro.cosme.e.silva@tecnico.ulisboa.pt}
\affiliation{GoLP/Instituto de Plasmas e Fus\~ao Nuclear, Instituto Superior T\'ecnico, 1049-001 Lisboa, Portugal}
\author{Hugo Terças}
\affiliation{GoLP/Instituto de Plasmas e Fus\~ao Nuclear, Instituto Superior T\'ecnico, 1049-001 Lisboa, Portugal}

\date{\today}

\begin{abstract}
The hydrodynamic behavior of charged carriers leads to nonlinear phenomena such as solitary waves and shocks, among others. As an application, such waves might be exploited on plasmonic devices either for modulation or signal propagation along graphene waveguides. We study the nature of nonlinear perturbations following an approach similar to Sagdeev potential analysis and also by performing the reductive perturbation method on the hydrodynamic description of graphene electrons, taking into consideration the effect of Bohm quantum potential and odd viscosity. Thus, deriving a dissipative Kadomtsev--Petviashvili--Burgers (KPB) equation for the bidimensional flow as well as its unidimensional limit in the form of Korteweg--de Vries--Burgers (KdVB). The stability analysis of these equations unveils the existence of unstable modes that can be excited and launched through graphene plasmonic devices.
\end{abstract}

\maketitle
\section{Introduction}
\label{intro}
The advent of graphene, and other two-dimensional materials, opened the way for remarkable and exciting physics and phenomena, particularly in the domains of charge transport and plasmonics where high mobility of electrons is required. Such areas are crucial to the development of next-generation devices that are compatible with integrated circuit technology \cite{Wang2017c}, such as transistors \cite{Murali2012}, quantum dots \cite{Chung2021}, radiation detectors and emitters or waveguides \cite{Sun2021Ultra-LowCommunication}. Indeed, the absence of gap in monolayer graphene, being problematic for digital devices, placed the focus of research on continuous wave applications, specially in the highly sought after THz range \cite{Concepts2020Graphene-BasedPlasmonics}. Much of the research on the THz problem in graphene devices take place within the hydrodynamic framework \cite{Tomadin2013, Concepts2020Graphene-BasedPlasmonics, man_2021, toshio_2022}, a feature that has been motivated by the recent theoretical and experimental works supporting the hydrodynamic regime of electrons in graphene \cite{Lucas2018, Narozhny2019ElectronicGraphene, Muller2009, Ku2020a, Sulpizio2019, Samaddar2021EvidenceMobilityb, Monch2022RatchetTransport, Krebs_2023}. Recent works in graphene hydrodynamics involve viscometry \cite{tomadin2014, Cook_2021}, electron-phonon coupling \cite{Huang2020, Levchenko2020TransportLiquids}, nonlocal resistivity \cite{Li2022NonlocalTransport, Torre2015a, Levitov2016ElectronGraphene}. Consequently, the investigation of hydrodynamic plasmonic instabilities has received a new breath within the different communities, namely through mechanisms such as Coulomb drag \cite{Ryzhii2022,Ryzhii2022a}, and Dyakonov--Shur and Ryzhii--Satou--Shur instabilities \cite{Satou2016a, Koseki2016, Cosme2020, Cosme2021}, the plasmonic boom instability \cite{Aizin2016Current-drivenNanostructures} and surface-plasmon polaritons \cite{Zolotovskii2018,Dong2021}. \par

A prominent advantage of the hydrodynamic formulation of graphene electrons in respect to the quantum kinetic formulations is the study of nonlinear phenomena: the hydrodynamic equation are more suitable for analytical and numerical methods \cite{Cosme2023TETHYS:Models}, despite ignoring some microscopic aspects of the momentum distributions in out-of-equilibrium situations. Tough the nonlinear effects in optical setups -- resorting to surface plasmon polaritons -- have been reported \cite{Ooi2017, Cox2019, Han2022}, the study of nonlinear dynamics in graphene plasmonic systems is still elusive. 

In this article, we explore the dynamics of nonlinear electron waves in graphene field-effect transistors (gFETs) within the framework of the hydrodynamic transport, achievable provided the following scale separation between the electron-electron free path ($\ell_{e-e}$), the inelastic free path ($\ell_{e-\text{imp}}$), and the system size ($L$) \cite{Lucas2018, Narozhny2019ElectronicGraphene}, 
\begin{equation}
    \ell_{e-e}\ll L\ll \ell_{e-\text{imp}},
    \label{eq:scale_separation}
\end{equation}
a condition that has been reported in several experimental papers \cite{Ku2020a,Sulpizio2019,Samaddar2021EvidenceMobilityb,Monch2022RatchetTransport}. Despite the apparent simplicity of treating the electron transport in graphene via hydrodynamic equations, there are three major points which set this models afar from regular hydrodynamics. The first one is the fact that the effective mass is {\it compressible and relativistic}, i.e. depends on the number density $n$ and on the flow speed $ v=\vert  \bf v \vert$ as \cite{Figueiredo2020}
\begin{equation}
\mathcal{M}=\frac{m^\star}{1-v^2/v_F^2},    
\end{equation}
where $m^\star = \hbar \sqrt{ \pi n}/v_F$ is the Drude mass \cite{Cosme2021,Chaves2017,Narozhny2019ElectronicGraphene}. Secondly, the existence of a non-diffusive, odd viscosity term, arising in two-dimensional systems with broken time-reversal symmetry \cite{Avron1998OddViscosity}, arising either from the presence of magnetic fields \cite{Narozhny2019MagnetohydrodynamicsViscosities,Chen2022ViscosityElectrons} or from anisotropy of the Fermi sphere \cite{Friedman2022HydrodynamicsGroup}. Finally, it has been recently introduced corrections in the form of a quantum (Bohm) potential \cite{Manfredi2001Self-consistentGas,Haas2011}, which can be obtained from a more complete quantum kinetic description \cite{Figueiredo2020}. All of these factors contribute to a peculiar competition between dispersion and nonlinearity in the graphene hydrodynamics, which has profound implications in the physics of the nonlinear, as we explain below.

In order to examine the several possibilities that lead to nonlinear waves, this work is organized in the following manner: in Section II, the base hydrodynamical model is presented; then, in section III, we derive the hamiltonian description of finite amplitude one-dimensional waves in the absence of viscosity. In section IV, we proceed to a perturbative method that allow us to deal with viscous terms and, finally, concluding remarks are presented in section V.

\section{Graphene hydrodynamic model}

We consider monolayer graphene in the field-effect transistor (FET) configuration, in which the gate $-$ located at a distance $d_0$ away from the graphene sheet $-$ effectively screens the Coulomb interaction between carriers \cite{Cosme2020}. In the fully degenerate limit, where the Fermi temperature is much higher than room temperature, the flow of electrons in gated graphene can be described by the following hydrodynamic model \cite{Cosme2021,Figueiredo2020}, comprising the continuity equation
\begin{equation}
\frac{\partial n}{\partial t} +\bm{\nabla}\!\!\cdot\! \left(n\mathbf{v} \right) = 0\label{eq:eulerequations1}    
\end{equation}
and the momentum conservation equation,
\begin{multline}
\frac{\partial }{\partial t}(nm^\star\vec{v}) + \bm{\nabla}\cdot\left(  nm^\star\vec{v}\otimes \vec{v}\right) +
\bm{\nabla}\cdot\bm\Pi + en\bm{\nabla}\phi +\\
+\frac{\mathcal{B}m^\star_0}{\sqrt{n_0}}\bm{\nabla}\cdot(\bm{\nabla}\!\otimes\!\bm{\nabla} \sqrt{n})=0.\label{eq:eulerequations2}%
\end{multline}
Here, $n$ and $\vec{v}=(u,v)$ are the density and the velocity fields, $\phi$ is the electrostatic potential, $\bm \Pi$ is the stress tensor. The last term in Eq. \eqref{eq:eulerequations2} is the quantum (Bohm) potential recently derived in \cite{Figueiredo2020}, with magnitude governed by the quantum mechanical coefficient  $\mathcal{B}=n_0\hbar^2/(32{m_0^\star}^2)$, given in terms of the equilibrium density $n_0$ and equilibrium mass $m_0^\star$. The stress tensor comprises both the Fermi quantum pressure, $p=\hbar v_F\sqrt{\pi n^3}/3$, and the shear and odd viscosities, $\eta_s$ and $\eta_o$, such that \cite{Narozhny2019MagnetohydrodynamicsViscosities}
\begin{equation}
    \bm{\nabla}\cdot\bm\Pi=\bm{\nabla}p-\eta_s\bm{\nabla}^2\vec{v}-\eta_o\bm{\nabla}^2(\vec{v}\times\hat{\vec{z}}).
\end{equation}
 Regarding the electric potential $\phi$, we assume the gradual-channel approximation \cite{Shur1990,Tomadin2013}, i.e., the electric potential is dominated by the gate potential, which effectively screens the Coulomb potential in the long wavelength limit $kd_0\ll 1$. Thus, the electrostatic force term in \eqref{eq:eulerequations2} reads 
\begin{equation}
\bm{\nabla}\phi=\frac{ed_0}{\varepsilon}\bm{\nabla}n+\frac{ed_0^3}{\varepsilon}\bm\nabla \nabla^2n,\label{eq:Ugate}
\end{equation}
where $d_0$ and $\varepsilon$ are the thickness and permittivity of the dielectric between the gate and graphene. The last term in Eq. \eqref{eq:Ugate} gives origin to dispersive corrections of order $\sim d_0/L$ to the plasmon velocity, and the associated effects were described in Ref. \cite{Svintsov2013}. Here, we are interested in long channel devices with strong gate shielding and hence we drop such corrections. Nonetheless, if desired, they can be easily incorporated as a normalization of the Bohm term.  

To consider infinitesimal perturbation along the channel, we linearize Eqs. \eqref{eq:eulerequations1} and \eqref{eq:eulerequations2} around the equilibrium, $n=n_0+n_1 e^{ikx-i\omega t}$, ${\bf v}={\bf v}_1 e^{ikx-i\omega t}$, which leads to the dispersion relation 
\begin{equation}
    \omega = Sk- i \frac{\nu_s}{2} k^2-\frac{ \mathcal{B}/n_0-\nu_o^2+(\nu_s/2)^2}{2 S}k^3 \label{eq:disprelation}
\end{equation}
with  $S=\sqrt{e^2d_0n_0/\varepsilon m^\star_0}$ being the plasmon group velocity and $\nu_{s,o}\equiv\eta_{s,o}/n_0m^\star_0$ the kinematic viscosity. Thus, it is clear that the inclusion of odd viscosity and quantum potential terms does not impact the attenuation of the plasma waves but rather enhances the nonlinearity of the spectrum even in the limit when $\nu_s\ll1$. The presence of this strong dispersion already hints for the possibility of solitonic solutions, as found in other quantum \cite{Ali2007} and relativistic plasmas \cite{Haas2015}, and as we show in what follows.

\section{Finite amplitude nonlinear waves}
\begin{figure}[t!]
    \includegraphics[width=.9\linewidth]{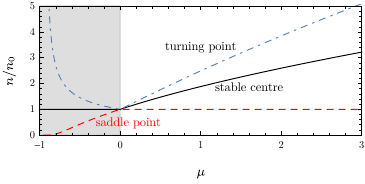}
    \caption{Transcritical bifurcation diagram of Eq. \eqref{eq:Equation_of_motion}, showing the position and nature of the fixed points and the turning point of the separatrix (blue dotted-dashed line). The bifurcation swaps the equilibrium (black solid line) and saddle (red dashed line) points.
    }
    \label{fig:bifurcation}
\end{figure}

\begin{figure}[ht!]
 \includegraphics[width=\linewidth]{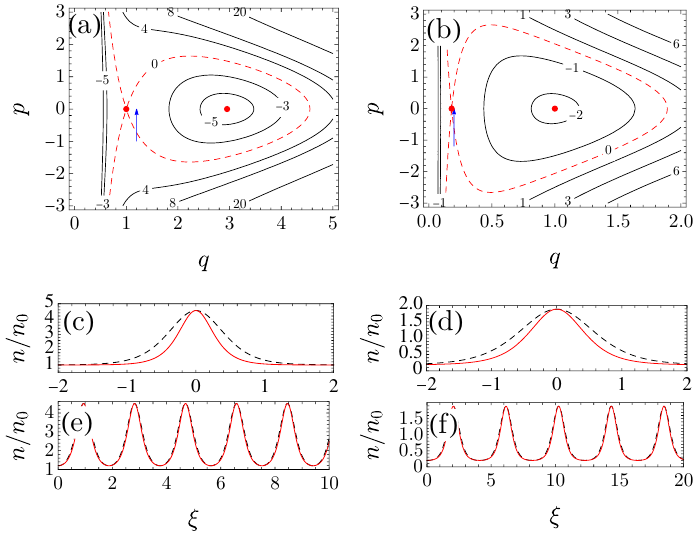}
\caption{(a-b) Phase space of the Hamiltonian \eqref{eq:hamiltonian} for $\mathcal{B}=1$, $S/v_F=2$ and $c/v_F=4$ (a) or $S/v_F=1.2$ (b). The fixed points $(n_0,0)$ and $(n_c,0)$ are marked by highlighted dots, and the initial conditions of the oscillatory numerical solutions are indicated by the arrow tip. Bounded orbits exist inside the separatrix (red dashed line). (c-f) Numerical solutions of orbits on the phase space. The solitary (c-d) and oscillatory (e-f) numerical solutions (red solid line) are compared against cnoidal  analytical expressions of the same amplitude and wavelength (black dashed line).} 
\label{fig:sagdeev_phase_space}
\end{figure}

We start by considering the inviscid limit of the model, $\eta_s=\eta_o=0$, in order to get a better understanding of the Bohm potential effects. Following the approach by Sagdeev \cite{Sagdeev1969NonlinearTheory,Sagdeev1988NonlinearChaos}, one can look  for one-dimensional traveling wave solutions of Eqs. \eqref{eq:eulerequations1} and \eqref{eq:eulerequations2} by introducing the variable $\xi=x-ct$, with $c$ denoting the wave velocity. This allows us to recast the hydrodynamic equations as
\begin{equation}
-c n'+(nu)'=0  \label{eq:cont_travel}  
\end{equation}
and
\begin{multline}
-c u'+\left(\frac{u^2}{4}+\frac{v_F^2}{2}\log n+2S^2\!\sqrt{\frac{n}{n_0}} \right)'+\\+\frac{\mathcal{B}}{n^{3/2}}\left(\!\sqrt{n}\right)'''=0.\label{eq:momentum_travel}
\end{multline}
Integrating the previous equations once, and imposing the asymptotic conditions $n=n_0$ and $u=0$ at infinity, one gets the equation of motion governing the density perturbations
\begin{multline}
    \mathcal{B}n''-\frac{3\mathcal{B}}{4n}\left(n'\right)^2
    +n_0^2 c^2\Bigg[1- \frac{n}{n_0}- \log \frac{n}{n_0}+\\
    +\frac{v_F^2}{2c^2}  \left(\frac{n^2}{n_0^2}-1\right)
    +\frac{4S^2}{5c^2} \left(\frac{n^{5/2}}{n_0^{5/2}}-1\right)\Bigg] =0.\label{eq:Equation_of_motion}
\end{multline}
The latter can then be multiplied by the quantity $n'/n^{3/2}$ and be integrated once more to reveal the first integral of motion, 
\begin{equation}
  \mathcal{J}= \mathcal{B}\frac{(n')^2}{2n^{3/2}}+\mathcal{V}(n),
\end{equation}
where $\mathcal{V}$ is the Sagdeev potential 
\begin{multline}
  \mathcal{V}(n) =\frac{2n_0^2 c^2}{\sqrt{n}}\Bigg[1-\frac{n}{n_0}+\log\frac{n}{n_0}+\\+\frac{v_F^2}{2c^2}\left(\frac{n^2}{3n_0^2}+1\right)+\frac{4S^2}{5c^2}\left(\frac{n^{5/2}}{4n_0^{5/2}}+1\right)
    \Bigg]
\end{multline}
Moreover, by defining the canonical vector $(q,p)=(n, \mathcal{B} n' /n^{3/2})$, it can be shown that the Hamiltonian 
\begin{equation}
    H(p,q)=\frac{p^2q^{3/2}}{2\mathcal{B}}+\mathcal{V}(q) \label{eq:hamiltonian}
\end{equation}
retrieves the equation of motion in \eqref{eq:Equation_of_motion}. Hence, given the form of the pseudo-potential $\mathcal{V}$, it is clear that the two-dimensional hamiltonian flow has two fixed points. One located at $(q,p)=(n_0,0)$ independently of the model parameters, besides mean density, and a second point wandering along the $q$ axis, $(q,p)=(n_c,0)$, where $n_c$ is a function of the model parameters. 
Such two points undergo a transcritical saddle-center bifurcation governed by the parameter $\mu=c^2/(S^2+v_F^2/2)-1$ as made evident by Fig.\ref{fig:bifurcation}, where we can see the fixed points colliding and swapping their nature. Furthermore, the position of the mobile fixed point is taken to be, up to first order,  $n_c/n_0 \approx c^2/(S^2+v_F^2/2)$.
The nature of this bifurcation ensures persistence of the stable center and saddle pair, and thus one can deduce the occurrence of nonlinear oscillations around the center point, provided that the hamiltonian level is lower than that of the saddle point, which defines the separatrix, i.e. $\Delta H=H-H_\text{separatrix}<0$. Figure \ref{fig:sagdeev_phase_space} illustrates the phase space of the system showing the stable region enclosed by the separatrix. It is evident that the system sustain nonlinear oscillations around the stable center point. Those are similar to cnoidal waves although, in fact, not elliptic functions, given the non-rational nature of the potential $\mathcal{V}(q)$. Yet, the soliton solutions, that live along the separatrix, are more narrow than the usual profile. Moreover, it is interesting to note that, while for $\mu>0$ the soliton amplitude scales as $A\sim\mu\sim c^2$; in the case of $\mu<0$, i.e. for the slow solitons, the amplitude strongly deviates from the linearity on $\mu$ (cf. Fig.\ref{fig:bifurcation}). Particularly in the limit of $c\to0$ we have $A\sim\mu^{-3/2}$.

The possibility to maintain and propagate solitary waves of substantial amplitude can be exploited to transmit pulsed signals along a graphene channel. However, to accommodate the dissipative effects on the analysis, we must resort to perturbative methods.

\section{Small amplitude nonlinear waves}

Although finite amplitude waves can be excited in the studied inviscid regime, the inclusion of viscous effects is central to a more faithful description of nonlinear waves in Dirac electronic fluids. In this section we will deal with the viscosities, both shear and odd, and with the dynamics in the transverse direction; to do so, we will now restrict ourselves to 
waves in the perturbative regime. 

The general procedure to implement the reductive perturbation method (RPM) \cite{Taniuti1968ReductiveI,Taniuti1969PerturbationI,Taniuti1983NonlinearWaves,Leblond2008TheApplications} starts with casting the model Eqs. \eqref{eq:eulerequations1} and \eqref{eq:eulerequations2} in a general quasilinear form 
\begin{multline}
    \Bigg[\frac{\partial }{\partial t}+\mat{A}^x\frac{\partial}{\partial x} +\mat{A}^y\frac{\partial}{\partial y}+\mat{K}\left(\frac{\partial^2}{\partial x^2}+\frac{\partial^2}{\partial y^2}\right)+\\ +
    \mat{H}^x\left(\frac{\partial^3}{\partial x^3}+\frac{\partial^2}{\partial x\partial y^2}\right) 
    \Bigg]\vec{U}=0,\label{eq:quasilinear}
\end{multline}
with the state vector $\vec{U}=(n,u,v)^T$ and the matrices:
\begin{equation}
\begin{aligned}
    \mat{A}^x= \begin{pmatrix}
  u & \phantom{-}n & 0 \\
  \frac{S^2}{n} & \phantom{-}\frac{u}{2} & 0 \\[.65em]
  0 & -\frac{v}{2} & u
 \end{pmatrix}, \quad&
     \mat{A}^y= \begin{pmatrix}
  v & 0 & \phantom{-}n \\
  0 & v & -\frac{u}{2} \\
  \frac{S^2}{n} & 0 & \phantom{-}\frac{v}{2}
 \end{pmatrix}\\
 \mat{K}= \begin{pmatrix}
  0 & 0 & 0 \\
  0 & -\nu_s & \nu_o \\
  0 & -\nu_o & -\nu_s
 \end{pmatrix}, \quad& 
 \mat{H}^x= \frac{\mathcal{B}}{2n_0^2}\begin{pmatrix}
  0 & 0 & 0 \\
  1 & 0 & 0 \\
  1 & 0 & 0
 \end{pmatrix}.   
\end{aligned}\label{eq:KPmatrices}
\end{equation}

\subsection{Dissipative Kadomtsev--Petviashvili
equation}

Performing a  Gardiner--Morikawa transformation \cite{Taniuti1983NonlinearWaves,Leblond2008TheApplications} on Eq.\eqref{eq:quasilinear}, with the introduction of the set of stretched variables
\begin{equation}
    \begin{aligned}
    &\xi=\epsilon^{1/2}(x-\lambda t),\\
    &\zeta=\epsilon y,\text{ and}\\
    &\tau=\epsilon^{3/2}t,
    \end{aligned}
\end{equation}
 where $\epsilon$ is a small perturbation parameter being also used for the expansion of the variables,
\begin{equation}
    \begin{aligned}
    &n=n_0+\epsilon n_1+\epsilon^2 n_2 +\cdots\\
    &u=\epsilon u_1+\epsilon^2 u_2+\cdots \\
    &v=\epsilon^{3/2} v_1+\epsilon^{5/2} v_2 +\cdots
    \end{aligned}
\end{equation}
as well as for the matrices, e.g. $\mat{A}^x=\mat{A}^x_0+\epsilon\mat{A}^x_1+\cdots$ and so on. 
The choice of the exponents of $\epsilon$ is such that propagation along $x$ is predominant, and the dispersion relation \eqref{eq:disprelation}  remains invariant. Additionally, in order to capture the effect of the dissipation of the second term on the RHS of Eq.\eqref{eq:disprelation}, the shear viscosity must also be scaled as $\nu_s=\epsilon^{1/2}\Tilde{\nu}_s$.

Moreover, in the context of the RPM, we can introduce the first order perturbation field $\varphi$ such that    
\begin{equation}
    \vec{U}_1=(\pm n_0/S,1,0)^T\varphi,\label{eq:phi_def}
\end{equation}
where the plus and minus sign refer to right or left propagating waves respectively,
and then we can derive a dissipative generalization of the Kadomtsev--Petviashvili (KP) \cite{Bartuccelli1985Kadomtsev-Petviashvili-BurgersWaves} equation (see Appendix \ref{app:KPB}) 
\begin{multline}\frac{\partial}{\partial \xi}
 \!\left(\frac{\partial \varphi}{\partial \tau }+\frac{3}{4}\varphi \frac{\partial\varphi}{\partial \xi }-\frac{\Tilde{\nu}_s}{2}
  \frac{\partial^2\varphi}{\partial \xi^2}
  \pm\frac{\mathcal{B}/n_0-\nu_o^2}{2S}\frac{\partial^3\varphi}{\partial \xi^3}
  \right)\pm\\\pm \frac{S}{2} \frac{\partial^2 \varphi}{\partial\zeta^2}=0.\label{eq:KP}
\end{multline} 
Akin to what is found in the literature for other quantum plasmas \cite{Ghosh2019NonlinearPlasma,Misra2015MultidimensionalPlasmas,Seadawy2017IonPlasma}. 

In the one-dimensional limit of \eqref{eq:KP} one retrieves a dissipative generalization of the well-known Kortweg--de Vries--Burgers (KdVB) equation 
\begin{equation}
    \frac{\partial \varphi}{\partial \tau }+\frac{3}{4}\varphi \frac{\partial\varphi}{\partial \xi }-\frac{\Tilde{\nu}_s}{2}
  \frac{\partial^2\varphi}{\partial \xi^2}
  \pm\frac{\mathcal{B}/n_0-\nu_o^2}{2S}\frac{\partial^3\varphi}{\partial \xi^3}=0
  \label{eq:KdVB}
\end{equation}
This equation admits both oscillatory and shock-type solutions \cite{Canosa1977TheEquation,Bona1985Travelling-waveEquation,Jeffrey1989ExactEquation}. While the travelling shocks may be valuable for signal propagation engineering schemes, the oscillatory modes may trigger instabilities that could, in future technological applications, be harnessed to excite radiative emission.  

\begin{figure}[t!]
    \centering
    \includegraphics[scale=1]{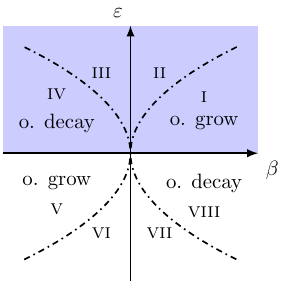}
    \caption{Parameter space regions with distinct qualitative behavior, bounded by $\varepsilon^4=16\beta^2$. Regions II, III, VI and VII only sustain bounded solutions along the heteroclinic orbit connecting the two fixed points. Whilst the remaining areas (labelled with o.) feature oscillatory solutions, either decaying or growing in time. Shaded region $\varepsilon\geq0$ indicating the achievable region of positive shear viscosity.}
    \label{fig:regions_analysis}
\end{figure}

Regarding the instance of unstable modes let us devote our attention to right-propagating solutions, setting $\chi=\xi-c\tau$ as independent variable, Eq.\eqref{eq:KdVB} can be cast to the adimensional form 
\begin{equation}
    -\varphi'+\frac{3}{4}\varphi\varphi'-\varepsilon\varphi''+\beta\varphi'''=0
    \label{eq:adimKdvB}
\end{equation}
with 
\begin{equation}
    \varepsilon\equiv \frac{\Tilde{\nu}_s}{2cL} \quad\text{and}\quad \beta \equiv \frac{\mathcal{B}/n_0-\nu_o^2}{2ScL^2},
\end{equation}
thus, the global stability and qualitative behavior of \eqref{eq:adimKdvB} can be analysed in terms of such parameters. 
Integration of \eqref{eq:adimKdvB} yields 
\begin{equation}
        -\varphi+\frac{3}{8}\varphi^2-\varepsilon\varphi'+\beta\varphi''=r\label{eq:KdVB2nd}
\end{equation}
with $r$ an integration constant, assuming that the perturbation field vanishes at infinity imposes $r=0$.
Thus, the flow linearization around the fixed points of \eqref{eq:KdVB2nd}, to wit, 
\begin{equation}
    (\varphi_-,\varphi'_-)=(0,0)\quad\text{and}\quad(\varphi_+,\varphi'_+)=\left(\frac{8}{3},0\right),
    \label{eq:fixed_points}
\end{equation}
yields the eigenvalues
\begin{equation}
    \lambda_{1,2}(\varphi_-)=\frac{\varepsilon\pm\sqrt{\varepsilon^2+4\beta}}{2\beta}\text{ and}
\end{equation}
\begin{equation}
    \lambda_{1,2}(\varphi_+)=\frac{\varepsilon\pm\sqrt{\varepsilon^2-4\beta}}{2\beta},
\end{equation}
therefore, the global behavior of the dynamical system can be categorized by the regions bounded by $\varepsilon^2\pm4\beta=0$ as illustrated on Fig.\ref{fig:regions_analysis}, while the behavior of the fixed points is listed on Tab. \ref{tab:tabela_phaseportrait}.

\begingroup
\squeezetable
\begin{table}[!ht]
    \centering
        \caption{Schematic diagrams and behavior of the phase portrait of Kortweg--de Vries--Burgers equation \eqref{eq:KdVB2nd} around the fixed points for the regions of parameters I through VIII.}
    \label{tab:tabela_phaseportrait}
\begin{ruledtabular}    
\begin{tabular*}{\linewidth}{c @{\extracolsep{\fill}}c@{\extracolsep{\fill}}c  c }
        \textbf{Region} & \textbf{Phase Portrait} & \textbf{Behavior} $\varphi_-$ & \textbf{Behavior} $\varphi_+$ \\[1ex]
 \hline
        I  & \raisebox{-.5\height}{\includegraphics[scale=0.4]{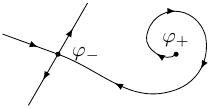}} & saddle & unstable spiral  \\
        II  & \raisebox{-.5\height}{\includegraphics[scale=0.4]{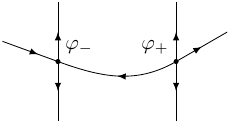}}
        & saddle & unstable node  \\
        III  & \raisebox{-.5\height}{\includegraphics[scale=0.4]{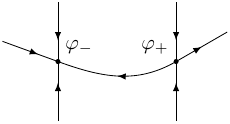}} & stable node & saddle  \\
        IV & \raisebox{-.5\height}{\includegraphics[scale=0.4]{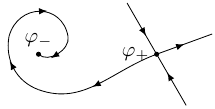}}  & stable spiral & saddle \\
        V  & \raisebox{-.5\height}{\includegraphics[scale=0.4]{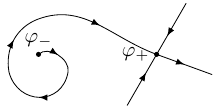}}  & unstable spiral & saddle \\
        VI   & \raisebox{-.5\height}{\includegraphics[scale=0.4]{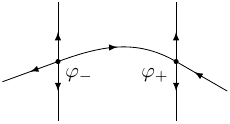}}  & unstable node & saddle \\
        VII& \raisebox{-.5\height}{\includegraphics[scale=0.4]{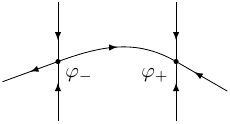}} & saddle  & stable node  \\
        VIII & \raisebox{-.5\height}{\includegraphics[scale=0.4]{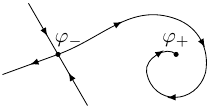}} & saddle & stable spiral  \\
\end{tabular*}
\end{ruledtabular}
\end{table}
\endgroup

Equation \eqref{eq:adimKdvB} can be seen as a combination of KdV and Burger's equations and,  indeed, its solutions present a crossover between the characteristic solutions of either KdV and Burger's, corresponding to the limits of negligible viscosity or dispersion, respectively.

\begin{figure*}[!ht]
    \centering
 \includegraphics[width=\linewidth]{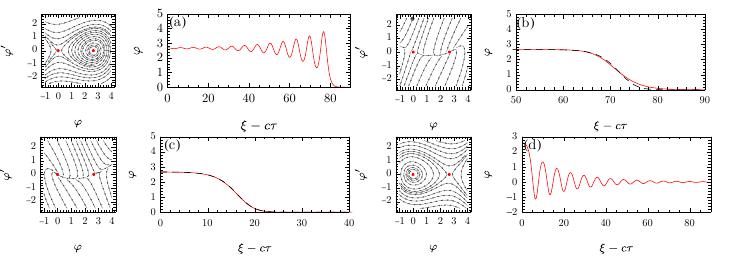} 
    \caption{Phase space (left, streamline plots) and numerical solutions (right, red solid line)of equation \eqref{eq:adimKdvB} for the positive viscosity regions, showing: (a) the growing oscillations $\varepsilon=0.1$, $\beta=1$, (b) shock propagation $\varepsilon=0.1$, $\beta=1$(c) idem $\varepsilon=5/\sqrt{6}$, $\beta=-1$, (d) decaying oscillations $\varepsilon=0.1$, $\beta=-1$. On panel (c) the analytical solution is superimposed (black dashed line) and on (b) a solution of the same form is also plotted for comparison. At the phase space plots the fixed points $(\varphi_\pm,\varphi_\pm')$ are highlighted (red dots).}
    \label{fig:solutions}
\end{figure*}

Notably, for regions I, IV, V and VIII, i.e. $|\beta|>\varepsilon^2/4$ the eigenvalues of one of the fixed points are complex conjugates, leading to stable (region IV) or unstable (region I) spirals. And, even tough the presence of viscosity breaks the homoclinic orbit, the system sustains oscillatory solutions, either decaying or increasing in time (cf. Fig.  \ref{fig:solutions}). For longer times, the self-growing modes  will
either collide with the hyperbolic point, reaching a local equilibrium or,  not being able to support themselves indefinitely and the nonlinear effects of the next order terms (that is, the response of $n_2$, $u_2$ and so on) will lead to the saturation or collapse of the wave. Nonetheless, even if short-lived, these modes can be used to trigger or reinforce other wave instabilities.

Further, for $|\beta|<\varepsilon^2/4$, as all eigenvalues are real, the only bounded solutions are those advancing on the heteroclinic orbit $\varphi'=(3\varphi^2 /8-\varphi)/\varepsilon$ connecting the fixed points. Consequently, the allowed solutions are sigmoid-like shock waves, similarly to the solutions of Burger's equation. In particular, for $|\beta|=6\varepsilon^2/25$ there is an analytical solution \cite{Feng2003ExactEquation,ABLOWITZ1979} in the form $\frac{8}{3}\left(1+e^{(\xi-c\tau-C_1)/\sqrt{6}}\right)^{-2},$ for other values of the ratio $\beta/\varepsilon^2$
the numerical solutions follow a similar profile.     

It has been argued in Ref. \cite{Kourakis2012} that only the shock solutions can have physical significance, the reasoning being that in the common scenarios -- often astrophysical ones -- the parameters of KdVB equation are not independent. However, in our system the coefficients are determined by a variety of physical parameters that can be set independently, viz. permittivity and thickness of the dielectric, Fermi level of the carriers, and both viscosities.     

\subsection{Modulational instability and nonlinear Schrödinger equation}

So far in this work, we considered only unmagnetized scenarios, where the quantum correction of the Bohm potential is crucial for the dispersive behavior of the waves. However, in the presence of a magnetic field, the onset of the cyclotron frequency ($\omega_c$) gap in the dispersion relation, causes the magnetic effects to dominate over those of the quantum potential. Therefore, we will now drop the Bohm contribution and focus on the nonlinear effects under a magnetic field.       

Introducing a weak magnetic field \cite{Cosme2021} to the model, with the addition of a $\omega_c\vec{v}\times\vec{\hat{z}}$ source term in Eq. \eqref{eq:eulerequations2}, leads to yet another nonlinear behavior -- the emergence of modulational instability. Indeed, from the hydrodynamic model written as 
\begin{equation}
    \left[\frac{\partial }{\partial t} +\mat{A}\frac{\partial }{\partial x} + \mat{K}\frac{\partial^2 }{\partial x^2}+\mat{B}\right]\vec{U}=0,\label{eq:diffusive_hydro}
\end{equation}
we can obtain the dispersion relation
\begin{equation}
    \omega^2=\omega_c^2+S^2k^2+\nu_o^2k^4-2\nu_o\omega_ck^2.\label{eq:magneto_plasmons}
\end{equation}
Then, following, once again, the prescription of the reductive perturbation method, now for the wave amplitude envelope  
\begin{equation}
    \vec{U}=\vec{U}_0+ \sum_{\substack{p=1\\|\ell|\leq p}}^\infty \epsilon^p\vec{U}^{(\ell)}_p e^{-i\ell(\omega t-kx)} 
\end{equation}
and defining the scalar $\psi$ from the first harmonic of the first order term and the appropriate right eigenvector of the differential operator of \eqref{eq:diffusive_hydro}, i.e.  
$\vec{U}_1^{(1)}\equiv \left(n_0,\pm\omega /k,i \left(k^2 \nu_o-\omega_c \right)/k\right)^T\psi$, one can derive (see the Appendix \ref{app:NLSE} for details) a nonlinear Schrödinger equation (NLSE) for the perturbation field, in the form
\begin{equation}
     i\frac{\partial \psi}{\partial \tau}+\frac{\omega''}{2}\frac{\partial^2 \psi}{\partial \xi^2}+Q\left|\psi\right|^2\psi=0, \label{eq:NLSE}
\end{equation}
where $\omega''\equiv\partial^2 \omega/\partial k^2$ and $Q$ the Kerr-like nonlinear term which can be cast as   
\begin{equation}
    Q = \frac{q(\omega,\omega_c,S,\nu_o)}{48n_0^2\omega\omega_c(4k^4 \nu_o^2-\omega_c^2)}
\end{equation}
with $q$ a somewhat complex polynomial given by:
\begin{widetext}
    \begin{multline}
        q(\omega,\omega_c,S,\nu_o) =-
     4 \omega_c (11 \omega^4 +9 \omega^2 \omega_c^2 - 
        8 \omega_c^4)+ 
     k^2 \Big[S^2 \omega_c (4 \omega^2 + 177 \omega_c^2) + 
        16 \nu_o (3 \omega^4 + 3 \omega^2 \omega_c^2 -20 \omega_c^4)\Big] +\\
       + 4 k^4 \Big[26 S^4 \omega_c + 
        3 S^2 \nu_o (4 \omega^2 - 57 \omega_c^2) + 
        3 \nu_o^2 \omega_c (7 \omega^2 + 96 \omega_c^2)\Big]
         -\\- 
     4 k^6 \nu_o \Big[24 S^4 - 231 S^2 \nu_o \omega_c + 
        8 \nu_o^2 (3 \omega^2 + 58 \omega_c^2)\Big] + 
     32 k^8 \nu_o^3 (43\nu_o\omega_c-15 S^2 ) -384 k^{10}\nu_o^5. 
    \end{multline}
\end{widetext}
where the wave vector $k$ is implicitly given by \eqref{eq:magneto_plasmons}. 

\begin{figure}
    \centering
    \includegraphics[width=\linewidth]{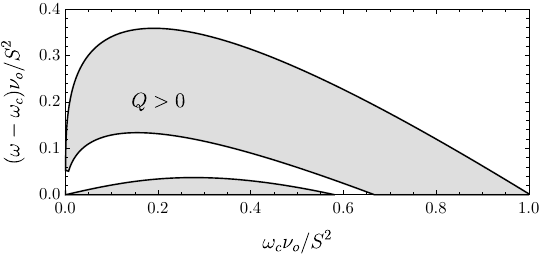}
    \caption{Region of instability for the nonlinear Schrödinger equation \eqref{eq:NLSE}. In the shaded regions, the positive Kerr term leads to a self-focusing (unstable) mode.}
    \label{fig:kerr_region}
\end{figure}

Such NLSE is known to foster the development of modulational instability \cite{Taniuti1983NonlinearWaves,Zakharov2009ModulationBeginning}. For the system to be unstable to modulations, it must comply with the Lighthill criterion $\omega''Q>0$, i.e. to be \emph{self-focusing} \cite{Kourakis2007}. Since for small $\nu_o$ the dispersion relation \eqref{eq:magneto_plasmons} ensures $\omega''>0$ the region of instability is determined by $Q$ alone; in fact, there is a region of parameters that leads to instability, as can be seen in Fig.~\ref{fig:kerr_region}. In that case, the spectral sidebands of a signal propagating in the system will grow, and the signal will increasingly modulate. Furthermore, in the limit of infinite wavelength of the modulation, the system can transmit a wave packet with an envelope governed by \eqref{eq:NLSE} -- i.e., a Peregrine soliton\cite{Peregrine1983WaterSolutions,Zakharov2013NonlinearInstability} -- similarly to the situation in optical media where this type of instability is well known and exploited.
 
\section{Conclusion}

The hydrodynamics of charged carriers on graphene has significant differences from regular fluid description of a two-dimensional electron gas, notably the local Drude mass, meaning that $m^\star(x,t)\propto\sqrt{n(x,t)}$, and the Bohm potential. As they enhance the dispersive nature of the flow, they favor the formation of nonlinear waves. We studied two classes of nonlinear waves: finite amplitude waves and perturbative waves. The former arises from a Sagdeev pseudo potential approach, while the latter stems from a reductive perturbative method that yields a generalized Kadomtsev--Petviashvili equation.\par

In the case of waves of general amplitude (although restricted to high Reynolds numbers), our finding reveal interesting properties: i) the formation of cnoidal-like waves that are not given in terms of elliptic functions, ii) the formation of solitons, propagating both above and below the group velocity of the linear plasmons. Remarkably, the latter violate the usual amplitude-velocity relation obtained for solitons withing the Kortweg--de Vries description \cite{zabusky1965}. In the perturbative scenario, the reduction of Kadomtsev--Petviashvili equation to the Kortweg--de Vries--Burgers equation exhibits regions of parameters for which oscillating shock waves are formed. In effect, the numerical solutions denote the transition between the viscosity dominated regime -- in which the solutions are pure (non oscillatory) shocks -- and the low viscosity case, for which nonlinear oscillations -- akin to the ones from Kortweg--de Vries equation -- superimposed with the shock profile are found. Furthermore, we extended the perturbative analysis of the magnetized case, retrieving a nonlinear Sch\r"odinger equation and showing the presence of modulational instability near the cyclotronic resonance. \par  

All the above points to the possible emergence of rather interesting nonlinear states, in particular propagating shock waves and solitons that can be exploited for plasmonic signal transmission along future graphene wave guides and circuitry. Moreover, the oscillatory unstable modes have the potential to drive the emission of radiation or to prompt even further unstable modes that we have not yet considered in our analysis, such that thermal instabilities and shock instabilities. Conversely, it is also foreseeable that the nonlinear effects described by our depiction will respond to external stimuli like radiation and temperature gradients, for instance. therefore, they may prove useful also for sensing applications.

\begin{acknowledgments}

The authors acknowledge Funda\c{c}\~{a}o para a Ci\^{e}ncia e a Tecnologia (FCT-Portugal) through Contract No. CEECIND/00401/2018, through the Project No. PTDC/FIS-OUT/3882/2020, the Exploratory Project No. UTA-EXPL/NPN/0038/2019, and through the Grant No. PD/BD/150415/2019.

\end{acknowledgments}

\appendix
\section{Reductive Perturbation Method}

\subsection{Dissipative Kadomtsev--Petviashvili}
\label{app:KPB}
The general procedure to implement the reductive perturbation method \cite{Taniuti1968ReductiveI,Taniuti1969PerturbationI,Taniuti1983NonlinearWaves,Leblond2008TheApplications} starts with casting the model Eqs. \eqref{eq:eulerequations1} and \eqref{eq:eulerequations2} in a general quasilinear form of Eq.\eqref{eq:quasilinear}
with the state vector $\vec{U}=(n,u,v)^T$ and the matrices defined at \eqref{eq:KPmatrices},
 Then, one performs a  Gardiner--Morikawa transformation \cite{Taniuti1983NonlinearWaves,Leblond2008TheApplications} introducing the set of stretched variables
\begin{equation}
    \begin{aligned}
    &\xi=\epsilon^{1/2}(x-\lambda t),\\
    &\zeta=\epsilon y,\text{ and}\\
    &\tau=\epsilon^{3/2}t,
    \end{aligned}
\end{equation}
 with $\epsilon$ being the small perturbation parameter being also used for the expansion of the variables,
as well as for the matrices, e.g. $\mat{A}^x=\mat{A}^x_0+\epsilon\mat{A}^x_1+\cdots$ and so on. 
Additionally, the shear viscosity must also be scaled as $\eta_s=\epsilon^{1/2}\Tilde{\eta}_s$ and so the matrix $\mat{K}$ must be split in $\mat{K}=\mat{K}_o+\epsilon^{1/2}\mat{K}_s$. 

Expanding equation \eqref{eq:quasilinear} as stated and equating each coefficient of the $\epsilon$ expansion to zero, yields, at the lowest order, that is $\epsilon^{3/2}$, 
\begin{equation}
    \left(\mat{A}_0^x-\lambda\right)\frac{\partial }{\partial \xi}\vec{U}_1=0,    
    \label{e3/2}
\end{equation}
while at the next order, $\epsilon^{2}$,
\begin{equation}
    \left(\mat{A}_0^x-\lambda\right)\frac{\partial }{\partial \xi}\vec{U}_{3/2}+\left(\mat{A}_0^y\frac{\partial }{\partial \zeta}+\mat{K}_o\frac{\partial^2 }{\partial \xi^2} \right)\vec{U}_1=0,    
    \label{e2}
\end{equation}
and for the order $\epsilon^{5/2}$, the last we will consider here, 
\begin{multline}
    \left(\mat{A}_0^x-\lambda\right)\frac{\partial }{\partial \xi}\vec{U}_{2}+\left(\mat{A}_0^y\frac{\partial }{\partial \zeta}+\mat{K}_o\frac{\partial^2 }{\partial \xi^2} \right)\vec{U}_{3/2}+\\+\left(\frac{\partial }{\partial \tau}+ \mat{A}_1^x\frac{\partial }{\partial \xi}+\mat{K}_{s}\frac{\partial^2 }{\partial \xi^2}+\mat{H}_0^x\frac{\partial^3 }{\partial \xi^3}\right)\vec{U}_1=0.    
    \label{e5/2}
\end{multline}
To simplify the previous equations, let the right and left eigenvectors of $\mat{A}_0^x$ associated with the eigenvalue $\lambda$ be defined as $\vec{L}_1$ and $\vec{R}_1$. Then, equation \eqref{e3/2} will be automatically satisfied if there is a scalar $\varphi_1$, which captures the time and spatial evolution, such that, 
\begin{equation}
    \vec{U}_1=\varphi\vec{R}_1. \label{eq:scalarfield}
\end{equation}
Moreover, multiplying equation \eqref{e5/2} by a left eigenvector of $A_0^x$ cancels the terms of $U_2$ leaving 
\begin{multline}
    \vec{L}_1\vec{R}_1\frac{\partial \varphi_1}{\partial \tau}+ \vec{L}_1\mat{A}_1^x\vec{R}_1\frac{\partial \varphi}{\partial \xi}+\vec{L}_1\mat{K}_s\vec{R}_1\frac{\partial^2 \varphi}{\partial \xi^2}+\\+\vec{L}_1\mat{H}_0^x\vec{R}_1\frac{\partial^3 \varphi}{\partial \xi^3}=-\vec{L}_1\left(\mat{A}_0^y\frac{\partial }{\partial \zeta}+\mat{K}_o\frac{\partial^2 }{\partial \xi^2} \right)\vec{U}_{3/2},\label{eq:generalKP}
\end{multline}
where we also made use of \eqref{eq:scalarfield}, and equation \eqref{e2} can now be used to simplify the RHS of this equation.

\subsection{Nonlinear Schrödinger equation}
\label{app:NLSE}
To derive the nonlinear Schrödinger equation (NLSE) from the fluid equations we will, once again, cast the system on its quasilinear form given by Eq. \eqref{eq:diffusive_hydro}
 we limit ourselves to 1D and discarded the quantum potential. Also, we only allow for linear source and diffusion terms; thus, $\mat{B}\equiv \mat{B}_0$ and  $\mat{K}\equiv\mat{K}_0$. For our hydrodynamic model we have: 
\begin{equation}
\begin{aligned}
    \mat{A}=
    \begin{pmatrix}u&\phantom{-}n&0\\
\frac{S^2}{n}&\phantom{-}\frac{u}{2}&0\\[0.65em]
0&-\frac{v}{2}&u\end{pmatrix},
\quad\mat{K}=
\begin{pmatrix}0&0&0\\
0&-\nu_s&\nu_o\\
0&-\nu_o&-\nu_s\end{pmatrix}\\
\quad\mat{B}=\begin{pmatrix}0&0&0\\
0&0&\omega_c\\
0&-\omega_c&0\end{pmatrix}.
\end{aligned}
\end{equation}
Expanding the wave amplitude envelope as
\begin{equation}
    \vec{U}=\vec{U}_0+ \sum_{p=1}^\infty \epsilon^p\vec{U}_p=\vec{U}_0+ \sum_{\substack{p=1\\|\ell|\leq p}}^\infty \epsilon^p\vec{U}^{(\ell)}_p e^{-i\ell(\omega t-kx)} 
\end{equation}
note that, since it $\vec{U}$ is a real function vector, $\vec{U}^{(-\ell)}_p\equiv {\vec{U}^{(\ell)}_p }^\ast$.
Moreover, let us expand the matrices as 
\begin{multline}
    \mat{A}=\mat{A}_0+ \sum_{p=1}^\infty\epsilon^p\mat{A}_p=\mat{A}_0+\epsilon \mat{A^\prime} [U_1]+\\+\epsilon^2\left(\mat{A^\prime} [U_2]+\mat{A^{\prime \prime}}[U_1U_1]\right) +\cdots
\end{multline}
where we used the following notation
\begin{equation}
    \mat{A^\prime} [X]\equiv\frac{\partial \mat{A}}{\partial u_k}X_k=\frac{\partial A_{ij}}{\partial u_k}X_k
\end{equation}
\begin{equation}
    \mat{A^{\prime \prime}}[XY]\equiv\frac{1}{2}\frac{\partial^2 \mat{A}}{\partial u_k\partial u_m}X_kY_m=\frac{1}{2}\frac{\partial^2 A_{ij}}{\partial u_k\partial u_m}X_kY_m
\end{equation}
Making use of the previous expansions and collecting the terms of \eqref{eq:diffusive_hydro} which are first order in $\epsilon$ we have 
\begin{equation}
    \sum_{\ell}\mat{W}^{(\ell)}_0\vec{U}_1 ^{(\ell)}e^{i\ell\vartheta}=0
\end{equation}
where we defined the matrix 
\begin{equation}
    \mat{W}^{(\ell)}_0=-i\ell\omega \mat{I}+i\ell k\mat{A}_0-\ell^2 k ^2\mat{K}_0+\mat{B}_0.
\end{equation}
The right eigenvector of the first mode is used to scale the first order perturbation, introducing the scalar $\psi$
\begin{equation}
 \mat{W}^{(1)}_0\vec{U}_1^{(1)}=0\iff \mat{W}^{(1)}_0  \vec{R}_1^{(1)}\psi=0
\end{equation}

Likewise, the second order terms lead to  
\begin{multline}
 \sum_{\ell}\left( \mat{W}_0^{(\ell)} \vec{U}_2^{(\ell)} +(\mat{A}_0-\lambda\mat{I}+2i\ell k\mat{K}_0)\frac{\partial \vec{U}_1 ^{(\ell)}}{\partial \xi}\right)e^{i\ell\vartheta}+\\+\sum_{\ell^{' }, m} i\ell' k\mat{A^\prime} [U^{(m)}_1]\vec{U}^{(\ell')}_1 e^{i(\ell'+m)\vartheta}=0
\end{multline}
which when equating the modes $\ell$ and $\ell'+m$ yield the following relations
\begin{equation}
    \vec{U}_2^{(0)}=\vec{R}_2^{(0)}\left|\psi\right|^2,\quad\vec{U}_2^{(1)}=\vec{R}_2^{(1)}
  \frac{\partial \psi}{\partial \xi}    \quad\text{and}\quad \vec{U}_2^{(2)}=\vec{R}_2^{(2)}{\psi}^2 
\end{equation}
where the $\vec{R}_2^{(\ell)}$ vectors are given by 
\begin{equation}
\begin{aligned}
    \vec{R}_2^{(0)}\equiv& -ik{\mat{W}_0^{(0)}}^{-1}\left(\mat{A'}[{R_1^{(1)}}^\ast]\vec{R}_1^{(1)}-\mat{A'}[R_1^{(1)}]{\vec{R}_1^{(1)}}^\ast\right)\\
      \vec{R}_2^{(1)}\equiv&-{\mat{W}_0^{(1)}}^{-1}(\mat{A}_0-\bm\lambda+2i\ell k\mat{K}_0)\vec{R}_1^{(1)}  \\
       \vec{R}_2^{(2)}\equiv&- ik{\mat{W}_0^{(2)}}^{-1}\mat{A^\prime} [R_1^{(1)}]\vec{R}_1^{(1)}
\end{aligned}
\end{equation}

Finally, from the third order we get 
\begin{multline}
 \sum_{\ell}\Bigg( \mat{W}_0^{(\ell)} \vec{U}_3^{(\ell)}+\frac{\partial \vec{U}_1^{(\ell)}}{\partial \tau}+\mat{K}_0\frac{\partial^2\vec{U}_1^{(\ell)}}{\partial \xi^2}+\\+(\mat{A}_0-\lambda\mat{I}+2i\ell k\mat{K}_0)\frac{\partial \vec{U}_2^{(\ell)}}{\partial \xi}\Bigg)e^{i\ell\vartheta}+\\+ \sum_{\ell', m}\left(ik \ell' \mat{A^\prime} [U^{(m)}_{1}]\vec{U}^{(\ell')}_2 +\mat{A^\prime} [U^{(m)}_{1}]\frac{\partial \vec{U}^{(\ell')}_1 }{\partial \xi}\right)e^{i(\ell'+m)\vartheta}+\\+
 \sum_{l, j, n} il k\mat{A^\prime} [U^{(j)}_2]\vec{U}^{(l)}_1 e^{i(l+j)\vartheta}+\\+il k\mat{A^{\prime \prime}}[U^{(j)}_{1}U^{(n)}_{1}]\vec{U}^{(l)}_1 e^{i(l+j+n)\vartheta}=0 
\end{multline}
collecting the first mode terms, one gets 
\begin{multline}
  \mat{W}_0^{(1)} \vec{U}_3^{(1)}+\vec{R}_1^{(1)}\frac{\partial \psi}{\partial \tau}+\mat{K}_0\vec{R}_1^{(1)}\frac{\partial^2\psi}{\partial \xi^2}+\\
  +(\mat{A}_0-\bm\lambda+2i k\mat{K}_0)\vec{R}_2^{(1)}\frac{\partial^2 \psi}{\partial \xi^2}+\\
  i2k \mat{A^\prime} [R^{(-1)}_{1}]\vec{R}^{(2)}_2\psi^* \psi^2+ik \mat{A^\prime} [R^{(0)}_{2}]\vec{R}^{(1)}_1 |\psi|^2\psi-\\
 -i k\mat{A^\prime} [R^{(2)}_2]\vec{R}_1^{(-1)}\psi^2\psi^* -i k\mat{A^{\prime \prime}}[R_1^{(1)}R_1^{(1)}]\vec{R}_1^{(-1)}\psi\psi\psi^*+\\+2i k\mat{A^{\prime \prime}}[R_1^{(1)}R_1^{(-1)}]\vec{R}_1^{(1)}\psi\psi^*\psi=0 
\end{multline}
multiplying this expression by a left eigenvector such that $\vec{L}_1^{(1)}\mat{W}_0^{(1)}=0$ and by $i$ we arrive to
\begin{multline}
     i\vec{L}_1^{(1)}\vec{R}_1^{(1)}\frac{\partial \psi}{\partial \tau}+\\+i\vec{L}_1^{(1)}\Big[\mat{K}_0\vec{R}_1^{(1)}+(\mat{A}_0-\lambda\mat{I}+2i k\mat{K}_0)\vec{R}_2^{(1)}\Big]\frac{\partial^2 \psi}{\partial \xi^2}-\\-k\vec{L}_1^{(1)}\Big(
  2\mat{A^\prime} [{R_1^{(1)}}^\ast] \vec{R}^{(2)}_2
 - \mat{A^\prime} [R^{(2)}_{2}]{\vec{R}_1^{(1)}}^\ast +\mat{
A^\prime} [R^{(2)}_{0}]\vec{R}_1^{(1)}+\\+
 2\mat{A^{\prime \prime}}[R_1^{(1)}{R_1^{(1)}}^\ast ]\vec{R}_1^{(1)}
 -\mat{A^{\prime \prime}}[R_1^{(1)}R_1^{(1)}]{\vec{R}_1^{(1)}}^\ast\Big)\left|\psi\right|^2\psi=0 
\end{multline}
which can be read as a nonlinear Schrödinger equation:

\begin{equation}
     i\frac{\partial \psi}{\partial \tau}+\frac{1}{2}\omega''\frac{\partial^2 \psi}{\partial \xi^2}+Q\left|\psi\right|^2\psi=0.
\end{equation}

To retrieve the instability criterion and growth rate one can resort to the ansatz $\psi=\left(\sqrt{P_0}+a(\xi,\tau)\right)e^{iQ P_0\tau}$ where
$a(\xi,\tau)=c_1e^{i(\kappa\xi-\Omega\tau)}+c_2e^{-i(\kappa\xi-\Omega\tau)}$. This leads to the dispersion relation 
\begin{equation}
    \Omega^2=\left(\frac{\omega''}{2}\right)^2\kappa^4-2\left(\frac{\omega''}{2}\right)QP_0\kappa^2 
\end{equation}
for the modulation frequency $\Omega$ and wave number $\kappa$. Evidently, for the frequency to be able to acquire an imaginary part $\omega''Q>0$ wich is our condition for instability. Whereas for $\omega''Q\leq0\Rightarrow\Omega\in\mathds{R}$. Still, even if this global criterion is satisfied, only the wave modes with $\kappa^2<2Q/\omega''$ do have imaginary frequency. It can also be shown that the maximum growth rate $\gamma_\text{max}\equiv\max_\kappa \Im \Omega(\kappa)=QP_0$; thus, proportional to the Kerr nonlinear term. 

\vspace*{\fill}
\bibliographystyle{apsrev4-2}
\bibliography{references}

\begin{thebibliography}{69}%
\makeatletter
\providecommand \@ifxundefined [1]{%
 \@ifx{#1\undefined}
}%
\providecommand \@ifnum [1]{%
 \ifnum #1\expandafter \@firstoftwo
 \else \expandafter \@secondoftwo
 \fi
}%
\providecommand \@ifx [1]{%
 \ifx #1\expandafter \@firstoftwo
 \else \expandafter \@secondoftwo
 \fi
}%
\providecommand \natexlab [1]{#1}%
\providecommand \enquote  [1]{``#1''}%
\providecommand \bibnamefont  [1]{#1}%
\providecommand \bibfnamefont [1]{#1}%
\providecommand \citenamefont [1]{#1}%
\providecommand \href@noop [0]{\@secondoftwo}%
\providecommand \href [0]{\begingroup \@sanitize@url \@href}%
\providecommand \@href[1]{\@@startlink{#1}\@@href}%
\providecommand \@@href[1]{\endgroup#1\@@endlink}%
\providecommand \@sanitize@url [0]{\catcode `\\12\catcode `\$12\catcode
  `\&12\catcode `\#12\catcode `\^12\catcode `\_12\catcode `\%12\relax}%
\providecommand \@@startlink[1]{}%
\providecommand \@@endlink[0]{}%
\providecommand \url  [0]{\begingroup\@sanitize@url \@url }%
\providecommand \@url [1]{\endgroup\@href {#1}{\urlprefix }}%
\providecommand \urlprefix  [0]{URL }%
\providecommand \Eprint [0]{\href }%
\providecommand \doibase [0]{https://doi.org/}%
\providecommand \selectlanguage [0]{\@gobble}%
\providecommand \bibinfo  [0]{\@secondoftwo}%
\providecommand \bibfield  [0]{\@secondoftwo}%
\providecommand \translation [1]{[#1]}%
\providecommand \BibitemOpen [0]{}%
\providecommand \bibitemStop [0]{}%
\providecommand \bibitemNoStop [0]{.\EOS\space}%
\providecommand \EOS [0]{\spacefactor3000\relax}%
\providecommand \BibitemShut  [1]{\csname bibitem#1\endcsname}%
\let\auto@bib@innerbib\@empty
\bibitem [{\citenamefont {Wang}\ \emph {et~al.}(2017)\citenamefont {Wang},
  \citenamefont {Duan},\ and\ \citenamefont {Duan}}]{Wang2017c}%
  \BibitemOpen
  \bibfield  {author} {\bibinfo {author} {\bibfnamefont {C.}~\bibnamefont
  {Wang}}, \bibinfo {author} {\bibfnamefont {X.}~\bibnamefont {Duan}},\ and\
  \bibinfo {author} {\bibfnamefont {X.}~\bibnamefont {Duan}},\ }in\ \href
  {https://doi.org/10.1017/9781316681619.010} {\emph {\bibinfo {booktitle} {2D
  Materials}}},\ \bibinfo {editor} {edited by\ \bibinfo {editor} {\bibfnamefont
  {P.}~\bibnamefont {Avouris}}, \bibinfo {editor} {\bibfnamefont {T.~F.}\
  \bibnamefont {Heinz}},\ and\ \bibinfo {editor} {\bibfnamefont
  {T.}~\bibnamefont {Low}}}\ (\bibinfo  {publisher} {Cambridge University
  Press},\ \bibinfo {address} {Cambridge},\ \bibinfo {year} {2017})\ pp.\
  \bibinfo {pages} {159--179}\BibitemShut {NoStop}%
\bibitem [{\citenamefont {Murali}(2012)}]{Murali2012}%
  \BibitemOpen
  \bibfield  {author} {\bibinfo {author} {\bibfnamefont {R.}~\bibnamefont
  {Murali}},\ }in\ \href {https://doi.org/10.1007/978-1-4614-0548-1{\_}3}
  {\emph {\bibinfo {booktitle} {Graphene Nanoelectronics}}},\ \bibinfo {editor}
  {edited by\ \bibinfo {editor} {\bibfnamefont {R.}~\bibnamefont {Murali}}}\
  (\bibinfo  {publisher} {Springer US},\ \bibinfo {address} {Boston, MA},\
  \bibinfo {year} {2012})\ pp.\ \bibinfo {pages} {51--91}\BibitemShut {NoStop}%
\bibitem [{\citenamefont {Chung}\ \emph {et~al.}(2021)\citenamefont {Chung},
  \citenamefont {Revia},\ and\ \citenamefont {Zhang}}]{Chung2021}%
  \BibitemOpen
  \bibfield  {author} {\bibinfo {author} {\bibfnamefont {S.}~\bibnamefont
  {Chung}}, \bibinfo {author} {\bibfnamefont {R.~A.}\ \bibnamefont {Revia}},\
  and\ \bibinfo {author} {\bibfnamefont {M.}~\bibnamefont {Zhang}},\ }\href
  {https://doi.org/10.1002/adma.201904362} {\bibfield  {journal} {\bibinfo
  {journal} {Advanced Materials}\ }\textbf {\bibinfo {volume} {33}},\ \bibinfo
  {pages} {1904362} (\bibinfo {year} {2021})}\BibitemShut {NoStop}%
\bibitem [{\citenamefont {Sun}\ \emph {et~al.}(2021)\citenamefont {Sun},
  \citenamefont {Huang}, \citenamefont {Wang}, \citenamefont {Lian},
  \citenamefont {Huang}, \citenamefont {Zhao},\ and\ \citenamefont
  {Zheng}}]{Sun2021Ultra-LowCommunication}%
  \BibitemOpen
  \bibfield  {author} {\bibinfo {author} {\bibfnamefont {L.}~\bibnamefont
  {Sun}}, \bibinfo {author} {\bibfnamefont {L.}~\bibnamefont {Huang}}, \bibinfo
  {author} {\bibfnamefont {Y.}~\bibnamefont {Wang}}, \bibinfo {author}
  {\bibfnamefont {Y.}~\bibnamefont {Lian}}, \bibinfo {author} {\bibfnamefont
  {G.}~\bibnamefont {Huang}}, \bibinfo {author} {\bibfnamefont
  {H.}~\bibnamefont {Zhao}},\ and\ \bibinfo {author} {\bibfnamefont
  {K.}~\bibnamefont {Zheng}},\ }\href
  {https://doi.org/10.1109/JPHOT.2021.3097334} {\bibfield  {journal} {\bibinfo
  {journal} {IEEE Photonics Journal}\ }\textbf {\bibinfo {volume} {13}},\
  \bibinfo {pages} {1} (\bibinfo {year} {2021})}\BibitemShut {NoStop}%
\bibitem [{\citenamefont {Mitin}\ \emph {et~al.}(2020)\citenamefont {Mitin},
  \citenamefont {Otsuji},\ and\ \citenamefont
  {Ryzhii}}]{Concepts2020Graphene-BasedPlasmonics}%
  \BibitemOpen
  \bibinfo {editor} {\bibfnamefont {V.}~\bibnamefont {Mitin}}, \bibinfo
  {editor} {\bibfnamefont {T.}~\bibnamefont {Otsuji}},\ and\ \bibinfo {editor}
  {\bibfnamefont {V.}~\bibnamefont {Ryzhii}},\ eds.,\ \href
  {https://doi.org/10.1201/9780429328398} {\emph {\bibinfo {title}
  {Graphene-Based Terahertz Electronics and Plasmonics}}},\ Vol.\ \bibinfo
  {volume} {148}\ (\bibinfo  {publisher} {Jenny Stanford Publishing},\ \bibinfo
  {year} {2020})\ pp.\ \bibinfo {pages} {148--162}\BibitemShut {NoStop}%
\bibitem [{\citenamefont {Tomadin}\ and\ \citenamefont
  {Polini}(2013)}]{Tomadin2013}%
  \BibitemOpen
  \bibfield  {author} {\bibinfo {author} {\bibfnamefont {A.}~\bibnamefont
  {Tomadin}}\ and\ \bibinfo {author} {\bibfnamefont {M.}~\bibnamefont
  {Polini}},\ }\href {https://doi.org/10.1103/PhysRevB.88.205426} {\bibfield
  {journal} {\bibinfo  {journal} {Physical Review B}\ }\textbf {\bibinfo
  {volume} {88}},\ \bibinfo {pages} {205426} (\bibinfo {year}
  {2013})}\BibitemShut {NoStop}%
\bibitem [{\citenamefont {Man}\ \emph {et~al.}(2021)\citenamefont {Man},
  \citenamefont {Xu}, \citenamefont {Xiao}, \citenamefont {Wen}, \citenamefont
  {Ding}, \citenamefont {Van~Duppen},\ and\ \citenamefont
  {Peeters}}]{man_2021}%
  \BibitemOpen
  \bibfield  {author} {\bibinfo {author} {\bibfnamefont {L.~F.}\ \bibnamefont
  {Man}}, \bibinfo {author} {\bibfnamefont {W.}~\bibnamefont {Xu}}, \bibinfo
  {author} {\bibfnamefont {Y.~M.}\ \bibnamefont {Xiao}}, \bibinfo {author}
  {\bibfnamefont {H.}~\bibnamefont {Wen}}, \bibinfo {author} {\bibfnamefont
  {L.}~\bibnamefont {Ding}}, \bibinfo {author} {\bibfnamefont {B.}~\bibnamefont
  {Van~Duppen}},\ and\ \bibinfo {author} {\bibfnamefont {F.~M.}\ \bibnamefont
  {Peeters}},\ }\href {https://doi.org/10.1103/PhysRevB.104.125420} {\bibfield
  {journal} {\bibinfo  {journal} {Phys. Rev. B}\ }\textbf {\bibinfo {volume}
  {104}},\ \bibinfo {pages} {125420} (\bibinfo {year} {2021})}\BibitemShut
  {NoStop}%
\bibitem [{\citenamefont {Toshio}\ and\ \citenamefont
  {Kawakami}(2022)}]{toshio_2022}%
  \BibitemOpen
  \bibfield  {author} {\bibinfo {author} {\bibfnamefont {R.}~\bibnamefont
  {Toshio}}\ and\ \bibinfo {author} {\bibfnamefont {N.}~\bibnamefont
  {Kawakami}},\ }\href {https://doi.org/10.1103/PhysRevB.106.L201301}
  {\bibfield  {journal} {\bibinfo  {journal} {Phys. Rev. B}\ }\textbf {\bibinfo
  {volume} {106}},\ \bibinfo {pages} {L201301} (\bibinfo {year}
  {2022})}\BibitemShut {NoStop}%
\bibitem [{\citenamefont {Lucas}\ and\ \citenamefont {Fong}(2018)}]{Lucas2018}%
  \BibitemOpen
  \bibfield  {author} {\bibinfo {author} {\bibfnamefont {A.}~\bibnamefont
  {Lucas}}\ and\ \bibinfo {author} {\bibfnamefont {K.~C.}\ \bibnamefont
  {Fong}},\ }\href {https://doi.org/10.1088/1361-648X/aaa274} {\bibfield
  {journal} {\bibinfo  {journal} {Journal of Physics: Condensed Matter}\
  }\textbf {\bibinfo {volume} {30}},\ \bibinfo {pages} {053001} (\bibinfo
  {year} {2018})}\BibitemShut {NoStop}%
\bibitem [{\citenamefont {Narozhny}(2019)}]{Narozhny2019ElectronicGraphene}%
  \BibitemOpen
  \bibfield  {author} {\bibinfo {author} {\bibfnamefont {B.~N.}\ \bibnamefont
  {Narozhny}},\ }\href {https://doi.org/10.1016/j.aop.2019.167979} {\bibfield
  {journal} {\bibinfo  {journal} {Annals of Physics}\ }\textbf {\bibinfo
  {volume} {411}},\ \bibinfo {pages} {167979} (\bibinfo {year}
  {2019})}\BibitemShut {NoStop}%
\bibitem [{\citenamefont {M{\"{u}}ller}\ \emph {et~al.}(2009)\citenamefont
  {M{\"{u}}ller}, \citenamefont {Schmalian},\ and\ \citenamefont
  {Fritz}}]{Muller2009}%
  \BibitemOpen
  \bibfield  {author} {\bibinfo {author} {\bibfnamefont {M.}~\bibnamefont
  {M{\"{u}}ller}}, \bibinfo {author} {\bibfnamefont {J.}~\bibnamefont
  {Schmalian}},\ and\ \bibinfo {author} {\bibfnamefont {L.}~\bibnamefont
  {Fritz}},\ }\href {https://doi.org/10.1103/PhysRevLett.103.025301} {\bibfield
   {journal} {\bibinfo  {journal} {Physical Review Letters}\ }\textbf {\bibinfo
  {volume} {103}},\ \bibinfo {pages} {25301} (\bibinfo {year}
  {2009})}\BibitemShut {NoStop}%
\bibitem [{\citenamefont {Ku}\ \emph {et~al.}(2020)\citenamefont {Ku},
  \citenamefont {Zhou}, \citenamefont {Li}, \citenamefont {Shin}, \citenamefont
  {Shi}, \citenamefont {Burch}, \citenamefont {Anderson}, \citenamefont
  {Pierce}, \citenamefont {Xie}, \citenamefont {Hamo}, \citenamefont {Vool},
  \citenamefont {Zhang}, \citenamefont {Casola}, \citenamefont {Taniguchi},
  \citenamefont {Watanabe}, \citenamefont {Fogler}, \citenamefont {Kim},
  \citenamefont {Yacoby},\ and\ \citenamefont {Walsworth}}]{Ku2020a}%
  \BibitemOpen
  \bibfield  {author} {\bibinfo {author} {\bibfnamefont {M.~J.~H.}\
  \bibnamefont {Ku}}, \bibinfo {author} {\bibfnamefont {T.~X.}\ \bibnamefont
  {Zhou}}, \bibinfo {author} {\bibfnamefont {Q.}~\bibnamefont {Li}}, \bibinfo
  {author} {\bibfnamefont {Y.~J.}\ \bibnamefont {Shin}}, \bibinfo {author}
  {\bibfnamefont {J.~K.}\ \bibnamefont {Shi}}, \bibinfo {author} {\bibfnamefont
  {C.}~\bibnamefont {Burch}}, \bibinfo {author} {\bibfnamefont {L.~E.}\
  \bibnamefont {Anderson}}, \bibinfo {author} {\bibfnamefont {A.~T.}\
  \bibnamefont {Pierce}}, \bibinfo {author} {\bibfnamefont {Y.}~\bibnamefont
  {Xie}}, \bibinfo {author} {\bibfnamefont {A.}~\bibnamefont {Hamo}}, \bibinfo
  {author} {\bibfnamefont {U.}~\bibnamefont {Vool}}, \bibinfo {author}
  {\bibfnamefont {H.}~\bibnamefont {Zhang}}, \bibinfo {author} {\bibfnamefont
  {F.}~\bibnamefont {Casola}}, \bibinfo {author} {\bibfnamefont
  {T.}~\bibnamefont {Taniguchi}}, \bibinfo {author} {\bibfnamefont
  {K.}~\bibnamefont {Watanabe}}, \bibinfo {author} {\bibfnamefont {M.~M.}\
  \bibnamefont {Fogler}}, \bibinfo {author} {\bibfnamefont {P.}~\bibnamefont
  {Kim}}, \bibinfo {author} {\bibfnamefont {A.}~\bibnamefont {Yacoby}},\ and\
  \bibinfo {author} {\bibfnamefont {R.~L.}\ \bibnamefont {Walsworth}},\ }\href
  {https://doi.org/10.1038/s41586-020-2507-2} {\bibfield  {journal} {\bibinfo
  {journal} {Nature}\ }\textbf {\bibinfo {volume} {583}},\ \bibinfo {pages}
  {537} (\bibinfo {year} {2020})}\BibitemShut {NoStop}%
\bibitem [{\citenamefont {Sulpizio}\ \emph {et~al.}(2019)\citenamefont
  {Sulpizio}, \citenamefont {Ella}, \citenamefont {Rozen}, \citenamefont
  {Birkbeck}, \citenamefont {Perello}, \citenamefont {Dutta}, \citenamefont
  {Ben-Shalom}, \citenamefont {Taniguchi}, \citenamefont {Watanabe},
  \citenamefont {Holder}, \citenamefont {Queiroz}, \citenamefont {Principi},
  \citenamefont {Stern}, \citenamefont {Scaffidi}, \citenamefont {Geim},\ and\
  \citenamefont {Ilani}}]{Sulpizio2019}%
  \BibitemOpen
  \bibfield  {author} {\bibinfo {author} {\bibfnamefont {J.~A.}\ \bibnamefont
  {Sulpizio}}, \bibinfo {author} {\bibfnamefont {L.}~\bibnamefont {Ella}},
  \bibinfo {author} {\bibfnamefont {A.}~\bibnamefont {Rozen}}, \bibinfo
  {author} {\bibfnamefont {J.}~\bibnamefont {Birkbeck}}, \bibinfo {author}
  {\bibfnamefont {D.~J.}\ \bibnamefont {Perello}}, \bibinfo {author}
  {\bibfnamefont {D.}~\bibnamefont {Dutta}}, \bibinfo {author} {\bibfnamefont
  {M.}~\bibnamefont {Ben-Shalom}}, \bibinfo {author} {\bibfnamefont
  {T.}~\bibnamefont {Taniguchi}}, \bibinfo {author} {\bibfnamefont
  {K.}~\bibnamefont {Watanabe}}, \bibinfo {author} {\bibfnamefont
  {T.}~\bibnamefont {Holder}}, \bibinfo {author} {\bibfnamefont
  {R.}~\bibnamefont {Queiroz}}, \bibinfo {author} {\bibfnamefont
  {A.}~\bibnamefont {Principi}}, \bibinfo {author} {\bibfnamefont
  {A.}~\bibnamefont {Stern}}, \bibinfo {author} {\bibfnamefont
  {T.}~\bibnamefont {Scaffidi}}, \bibinfo {author} {\bibfnamefont {A.~K.}\
  \bibnamefont {Geim}},\ and\ \bibinfo {author} {\bibfnamefont
  {S.}~\bibnamefont {Ilani}},\ }\href
  {https://doi.org/10.1038/s41586-019-1788-9} {\bibfield  {journal} {\bibinfo
  {journal} {Nature}\ }\textbf {\bibinfo {volume} {576}},\ \bibinfo {pages}
  {75} (\bibinfo {year} {2019})}\BibitemShut {NoStop}%
\bibitem [{\citenamefont {Samaddar}\ \emph {et~al.}(2021)\citenamefont
  {Samaddar}, \citenamefont {Strasdas}, \citenamefont {Jan{\ss}en},
  \citenamefont {Just}, \citenamefont {Johnsen}, \citenamefont {Wang},
  \citenamefont {Uzlu}, \citenamefont {Li}, \citenamefont {Neumaier},
  \citenamefont {Liebmann},\ and\ \citenamefont
  {Morgenstern}}]{Samaddar2021EvidenceMobilityb}%
  \BibitemOpen
  \bibfield  {author} {\bibinfo {author} {\bibfnamefont {S.}~\bibnamefont
  {Samaddar}}, \bibinfo {author} {\bibfnamefont {J.}~\bibnamefont {Strasdas}},
  \bibinfo {author} {\bibfnamefont {K.}~\bibnamefont {Jan{\ss}en}}, \bibinfo
  {author} {\bibfnamefont {S.}~\bibnamefont {Just}}, \bibinfo {author}
  {\bibfnamefont {T.}~\bibnamefont {Johnsen}}, \bibinfo {author} {\bibfnamefont
  {Z.}~\bibnamefont {Wang}}, \bibinfo {author} {\bibfnamefont {B.}~\bibnamefont
  {Uzlu}}, \bibinfo {author} {\bibfnamefont {S.}~\bibnamefont {Li}}, \bibinfo
  {author} {\bibfnamefont {D.}~\bibnamefont {Neumaier}}, \bibinfo {author}
  {\bibfnamefont {M.}~\bibnamefont {Liebmann}},\ and\ \bibinfo {author}
  {\bibfnamefont {M.}~\bibnamefont {Morgenstern}},\ }\href
  {https://doi.org/10.1021/acs.nanolett.1c01145} {\bibfield  {journal}
  {\bibinfo  {journal} {Nano Letters}\ }\textbf {\bibinfo {volume} {21}},\
  \bibinfo {pages} {9365} (\bibinfo {year} {2021})}\BibitemShut {NoStop}%
\bibitem [{\citenamefont {M{\"{o}}nch}\ \emph {et~al.}(2022)\citenamefont
  {M{\"{o}}nch}, \citenamefont {Potashin}, \citenamefont {Lindner},
  \citenamefont {Yahniuk}, \citenamefont {Golub}, \citenamefont {Kachorovskii},
  \citenamefont {Bel'kov}, \citenamefont {Huber}, \citenamefont {Watanabe},
  \citenamefont {Taniguchi}, \citenamefont {Eroms}, \citenamefont {Weiss},\
  and\ \citenamefont {Ganichev}}]{Monch2022RatchetTransport}%
  \BibitemOpen
  \bibfield  {author} {\bibinfo {author} {\bibfnamefont {E.}~\bibnamefont
  {M{\"{o}}nch}}, \bibinfo {author} {\bibfnamefont {S.~O.}\ \bibnamefont
  {Potashin}}, \bibinfo {author} {\bibfnamefont {K.}~\bibnamefont {Lindner}},
  \bibinfo {author} {\bibfnamefont {I.}~\bibnamefont {Yahniuk}}, \bibinfo
  {author} {\bibfnamefont {L.~E.}\ \bibnamefont {Golub}}, \bibinfo {author}
  {\bibfnamefont {V.~Y.}\ \bibnamefont {Kachorovskii}}, \bibinfo {author}
  {\bibfnamefont {V.~V.}\ \bibnamefont {Bel'kov}}, \bibinfo {author}
  {\bibfnamefont {R.}~\bibnamefont {Huber}}, \bibinfo {author} {\bibfnamefont
  {K.}~\bibnamefont {Watanabe}}, \bibinfo {author} {\bibfnamefont
  {T.}~\bibnamefont {Taniguchi}}, \bibinfo {author} {\bibfnamefont
  {J.}~\bibnamefont {Eroms}}, \bibinfo {author} {\bibfnamefont
  {D.}~\bibnamefont {Weiss}},\ and\ \bibinfo {author} {\bibfnamefont {S.~D.}\
  \bibnamefont {Ganichev}},\ }\href
  {https://doi.org/10.1103/PhysRevB.105.045404} {\bibfield  {journal} {\bibinfo
   {journal} {Physical Review B}\ }\textbf {\bibinfo {volume} {105}},\ \bibinfo
  {pages} {045404} (\bibinfo {year} {2022})}\BibitemShut {NoStop}%
\bibitem [{\citenamefont {Krebs}\ \emph {et~al.}(2023)\citenamefont {Krebs},
  \citenamefont {Behn}, \citenamefont {Li}, \citenamefont {Smith},
  \citenamefont {Watanabe}, \citenamefont {Taniguchi}, \citenamefont
  {Levchenko},\ and\ \citenamefont {Brar}}]{Krebs_2023}%
  \BibitemOpen
  \bibfield  {author} {\bibinfo {author} {\bibfnamefont {Z.~J.}\ \bibnamefont
  {Krebs}}, \bibinfo {author} {\bibfnamefont {W.~A.}\ \bibnamefont {Behn}},
  \bibinfo {author} {\bibfnamefont {S.}~\bibnamefont {Li}}, \bibinfo {author}
  {\bibfnamefont {K.~J.}\ \bibnamefont {Smith}}, \bibinfo {author}
  {\bibfnamefont {K.}~\bibnamefont {Watanabe}}, \bibinfo {author}
  {\bibfnamefont {T.}~\bibnamefont {Taniguchi}}, \bibinfo {author}
  {\bibfnamefont {A.}~\bibnamefont {Levchenko}},\ and\ \bibinfo {author}
  {\bibfnamefont {V.~W.}\ \bibnamefont {Brar}},\ }\href
  {https://doi.org/10.1126/science.abm6073} {\bibfield  {journal} {\bibinfo
  {journal} {Science}\ }\textbf {\bibinfo {volume} {379}},\ \bibinfo {pages}
  {671} (\bibinfo {year} {2023})}\BibitemShut {NoStop}%
\bibitem [{\citenamefont {Tomadin}\ \emph {et~al.}(2014)\citenamefont
  {Tomadin}, \citenamefont {Vignale},\ and\ \citenamefont
  {Polini}}]{tomadin2014}%
  \BibitemOpen
  \bibfield  {author} {\bibinfo {author} {\bibfnamefont {A.}~\bibnamefont
  {Tomadin}}, \bibinfo {author} {\bibfnamefont {G.}~\bibnamefont {Vignale}},\
  and\ \bibinfo {author} {\bibfnamefont {M.}~\bibnamefont {Polini}},\ }\href
  {https://doi.org/10.1103/PhysRevLett.113.235901} {\bibfield  {journal}
  {\bibinfo  {journal} {Phys. Rev. Lett.}\ }\textbf {\bibinfo {volume} {113}},\
  \bibinfo {pages} {235901} (\bibinfo {year} {2014})}\BibitemShut {NoStop}%
\bibitem [{\citenamefont {Cook}\ and\ \citenamefont {Lucas}(2021)}]{Cook_2021}%
  \BibitemOpen
  \bibfield  {author} {\bibinfo {author} {\bibfnamefont {C.~Q.}\ \bibnamefont
  {Cook}}\ and\ \bibinfo {author} {\bibfnamefont {A.}~\bibnamefont {Lucas}},\
  }\href {https://doi.org/10.1103/PhysRevLett.127.176603} {\bibfield  {journal}
  {\bibinfo  {journal} {Phys. Rev. Lett.}\ }\textbf {\bibinfo {volume} {127}},\
  \bibinfo {pages} {176603} (\bibinfo {year} {2021})}\BibitemShut {NoStop}%
\bibitem [{\citenamefont {Huang}\ and\ \citenamefont
  {Lucas}(2021)}]{Huang2020}%
  \BibitemOpen
  \bibfield  {author} {\bibinfo {author} {\bibfnamefont {X.}~\bibnamefont
  {Huang}}\ and\ \bibinfo {author} {\bibfnamefont {A.}~\bibnamefont {Lucas}},\
  }\href {https://doi.org/10.1103/PhysRevB.103.155128} {\bibfield  {journal}
  {\bibinfo  {journal} {Physical Review B}\ }\textbf {\bibinfo {volume}
  {103}},\ \bibinfo {pages} {155128} (\bibinfo {year} {2021})},\ \Eprint
  {https://arxiv.org/abs/2009.10084} {2009.10084} \BibitemShut {NoStop}%
\bibitem [{\citenamefont {Levchenko}\ and\ \citenamefont
  {Schmalian}(2020)}]{Levchenko2020TransportLiquids}%
  \BibitemOpen
  \bibfield  {author} {\bibinfo {author} {\bibfnamefont {A.}~\bibnamefont
  {Levchenko}}\ and\ \bibinfo {author} {\bibfnamefont {J.}~\bibnamefont
  {Schmalian}},\ }\href {https://doi.org/10.1016/j.aop.2020.168218} {\bibfield
  {journal} {\bibinfo  {journal} {Annals of Physics}\ }\textbf {\bibinfo
  {volume} {419}},\ \bibinfo {pages} {168218} (\bibinfo {year}
  {2020})}\BibitemShut {NoStop}%
\bibitem [{\citenamefont {Li}\ and\ \citenamefont
  {Levchenko}(2022)}]{Li2022NonlocalTransport}%
  \BibitemOpen
  \bibfield  {author} {\bibinfo {author} {\bibfnamefont {S.}~\bibnamefont
  {Li}}\ and\ \bibinfo {author} {\bibfnamefont {A.}~\bibnamefont {Levchenko}},\
  }\href {https://doi.org/10.1103/PhysRevB.105.L241405} {\bibfield  {journal}
  {\bibinfo  {journal} {Physical Review B}\ }\textbf {\bibinfo {volume}
  {105}},\ \bibinfo {pages} {L241405} (\bibinfo {year} {2022})}\BibitemShut
  {NoStop}%
\bibitem [{\citenamefont {Torre}\ \emph {et~al.}(2015)\citenamefont {Torre},
  \citenamefont {Tomadin}, \citenamefont {Geim},\ and\ \citenamefont
  {Polini}}]{Torre2015a}%
  \BibitemOpen
  \bibfield  {author} {\bibinfo {author} {\bibfnamefont {I.}~\bibnamefont
  {Torre}}, \bibinfo {author} {\bibfnamefont {A.}~\bibnamefont {Tomadin}},
  \bibinfo {author} {\bibfnamefont {A.~K.}\ \bibnamefont {Geim}},\ and\
  \bibinfo {author} {\bibfnamefont {M.}~\bibnamefont {Polini}},\ }\href
  {https://doi.org/10.1103/PhysRevB.92.165433} {\bibfield  {journal} {\bibinfo
  {journal} {Physical Review B - Condensed Matter and Materials Physics}\
  }\textbf {\bibinfo {volume} {92}},\ \bibinfo {pages} {1} (\bibinfo {year}
  {2015})},\ \Eprint {https://arxiv.org/abs/1508.00363} {1508.00363}
  \BibitemShut {NoStop}%
\bibitem [{\citenamefont {Levitov}\ and\ \citenamefont
  {Falkovich}(2016)}]{Levitov2016ElectronGraphene}%
  \BibitemOpen
  \bibfield  {author} {\bibinfo {author} {\bibfnamefont {L.}~\bibnamefont
  {Levitov}}\ and\ \bibinfo {author} {\bibfnamefont {G.}~\bibnamefont
  {Falkovich}},\ }\href {https://doi.org/10.1038/nphys3667} {\bibfield
  {journal} {\bibinfo  {journal} {Nature Physics}\ }\textbf {\bibinfo {volume}
  {12}},\ \bibinfo {pages} {672} (\bibinfo {year} {2016})}\BibitemShut
  {NoStop}%
\bibitem [{\citenamefont {Ryzhii}\ \emph
  {et~al.}(2022{\natexlab{a}})\citenamefont {Ryzhii}, \citenamefont {Ryzhii},
  \citenamefont {Satou}, \citenamefont {Mitin}, \citenamefont {Shur},\ and\
  \citenamefont {Otsuji}}]{Ryzhii2022}%
  \BibitemOpen
  \bibfield  {author} {\bibinfo {author} {\bibfnamefont {V.}~\bibnamefont
  {Ryzhii}}, \bibinfo {author} {\bibfnamefont {M.}~\bibnamefont {Ryzhii}},
  \bibinfo {author} {\bibfnamefont {A.}~\bibnamefont {Satou}}, \bibinfo
  {author} {\bibfnamefont {V.}~\bibnamefont {Mitin}}, \bibinfo {author}
  {\bibfnamefont {M.}~\bibnamefont {Shur}},\ and\ \bibinfo {author}
  {\bibfnamefont {T.}~\bibnamefont {Otsuji}},\ }in\ \href
  {https://doi.org/10.1109/irmmw-thz50927.2022.9966853} {\emph {\bibinfo
  {booktitle} {2022 47th International Conference on Infrared, Millimeter and
  Terahertz Waves ({IRMMW}-{THz})}}}\ (\bibinfo  {publisher} {{IEEE}},\
  \bibinfo {year} {2022})\BibitemShut {NoStop}%
\bibitem [{\citenamefont {Ryzhii}\ \emph
  {et~al.}(2022{\natexlab{b}})\citenamefont {Ryzhii}, \citenamefont {Ryzhii},
  \citenamefont {Otsuji}, \citenamefont {Mitin},\ and\ \citenamefont
  {Shur}}]{Ryzhii2022a}%
  \BibitemOpen
  \bibfield  {author} {\bibinfo {author} {\bibfnamefont {M.}~\bibnamefont
  {Ryzhii}}, \bibinfo {author} {\bibfnamefont {V.}~\bibnamefont {Ryzhii}},
  \bibinfo {author} {\bibfnamefont {T.}~\bibnamefont {Otsuji}}, \bibinfo
  {author} {\bibfnamefont {V.}~\bibnamefont {Mitin}},\ and\ \bibinfo {author}
  {\bibfnamefont {M.~S.}\ \bibnamefont {Shur}},\ }\href
  {https://doi.org/10.1063/5.0087678} {\bibfield  {journal} {\bibinfo
  {journal} {Applied Physics Letters}\ }\textbf {\bibinfo {volume} {120}},\
  \bibinfo {pages} {111102} (\bibinfo {year} {2022}{\natexlab{b}})}\BibitemShut
  {NoStop}%
\bibitem [{\citenamefont {Satou}\ \emph {et~al.}(2016)\citenamefont {Satou},
  \citenamefont {Koseki}, \citenamefont {Watanabe}, \citenamefont {Popov},
  \citenamefont {Ryzhii},\ and\ \citenamefont {Otsuji}}]{Satou2016a}%
  \BibitemOpen
  \bibfield  {author} {\bibinfo {author} {\bibfnamefont {A.}~\bibnamefont
  {Satou}}, \bibinfo {author} {\bibfnamefont {Y.}~\bibnamefont {Koseki}},
  \bibinfo {author} {\bibfnamefont {T.}~\bibnamefont {Watanabe}}, \bibinfo
  {author} {\bibfnamefont {V.~V.}\ \bibnamefont {Popov}}, \bibinfo {author}
  {\bibfnamefont {V.}~\bibnamefont {Ryzhii}},\ and\ \bibinfo {author}
  {\bibfnamefont {T.}~\bibnamefont {Otsuji}}\ }(\bibinfo {year} {2016})\ p.\
  \bibinfo {pages} {98560F}\BibitemShut {NoStop}%
\bibitem [{\citenamefont {Koseki}\ \emph {et~al.}(2016)\citenamefont {Koseki},
  \citenamefont {Ryzhii}, \citenamefont {Otsuji}, \citenamefont {Popov},\ and\
  \citenamefont {Satou}}]{Koseki2016}%
  \BibitemOpen
  \bibfield  {author} {\bibinfo {author} {\bibfnamefont {Y.}~\bibnamefont
  {Koseki}}, \bibinfo {author} {\bibfnamefont {V.}~\bibnamefont {Ryzhii}},
  \bibinfo {author} {\bibfnamefont {T.}~\bibnamefont {Otsuji}}, \bibinfo
  {author} {\bibfnamefont {V.~V.}\ \bibnamefont {Popov}},\ and\ \bibinfo
  {author} {\bibfnamefont {A.}~\bibnamefont {Satou}},\ }\href
  {https://doi.org/10.1103/PhysRevB.93.245408} {\bibfield  {journal} {\bibinfo
  {journal} {Physical Review B}\ }\textbf {\bibinfo {volume} {93}},\ \bibinfo
  {pages} {245408} (\bibinfo {year} {2016})},\ \Eprint
  {https://arxiv.org/abs/1601.08108} {1601.08108} \BibitemShut {NoStop}%
\bibitem [{\citenamefont {Cosme}\ and\ \citenamefont
  {Ter{\c{c}}as}(2020)}]{Cosme2020}%
  \BibitemOpen
  \bibfield  {author} {\bibinfo {author} {\bibfnamefont {P.}~\bibnamefont
  {Cosme}}\ and\ \bibinfo {author} {\bibfnamefont {H.}~\bibnamefont
  {Ter{\c{c}}as}},\ }\href {https://doi.org/10.1021/acsphotonics.0c00313}
  {\bibfield  {journal} {\bibinfo  {journal} {ACS Photonics}\ }\textbf
  {\bibinfo {volume} {7}},\ \bibinfo {pages} {1375} (\bibinfo {year} {2020})},\
  \Eprint {https://arxiv.org/abs/1905.05536} {1905.05536} \BibitemShut
  {NoStop}%
\bibitem [{\citenamefont {Cosme}\ and\ \citenamefont
  {Ter{\c{c}}as}(2021)}]{Cosme2021}%
  \BibitemOpen
  \bibfield  {author} {\bibinfo {author} {\bibfnamefont {P.}~\bibnamefont
  {Cosme}}\ and\ \bibinfo {author} {\bibfnamefont {H.}~\bibnamefont
  {Ter{\c{c}}as}},\ }\href {https://doi.org/10.1063/5.0045444} {\bibfield
  {journal} {\bibinfo  {journal} {Applied Physics Letters}\ }\textbf {\bibinfo
  {volume} {118}},\ \bibinfo {pages} {131109} (\bibinfo {year}
  {2021})}\BibitemShut {NoStop}%
\bibitem [{\citenamefont {Aizin}\ \emph {et~al.}(2016)\citenamefont {Aizin},
  \citenamefont {Mikalopas},\ and\ \citenamefont
  {Shur}}]{Aizin2016Current-drivenNanostructures}%
  \BibitemOpen
  \bibfield  {author} {\bibinfo {author} {\bibfnamefont {G.~R.}\ \bibnamefont
  {Aizin}}, \bibinfo {author} {\bibfnamefont {J.}~\bibnamefont {Mikalopas}},\
  and\ \bibinfo {author} {\bibfnamefont {M.}~\bibnamefont {Shur}},\ }\href
  {https://doi.org/10.1103/PhysRevB.93.195315} {\bibfield  {journal} {\bibinfo
  {journal} {Physical Review B}\ }\textbf {\bibinfo {volume} {93}},\ \bibinfo
  {pages} {195315} (\bibinfo {year} {2016})}\BibitemShut {NoStop}%
\bibitem [{\citenamefont {Zolotovskii}\ \emph {et~al.}(2018)\citenamefont
  {Zolotovskii}, \citenamefont {Dadoenkova}, \citenamefont {Moiseev},
  \citenamefont {Kadochkin}, \citenamefont {Svetukhin},\ and\ \citenamefont
  {Fotiadi}}]{Zolotovskii2018}%
  \BibitemOpen
  \bibfield  {author} {\bibinfo {author} {\bibfnamefont {I.~O.}\ \bibnamefont
  {Zolotovskii}}, \bibinfo {author} {\bibfnamefont {Y.~S.}\ \bibnamefont
  {Dadoenkova}}, \bibinfo {author} {\bibfnamefont {S.~G.}\ \bibnamefont
  {Moiseev}}, \bibinfo {author} {\bibfnamefont {A.~S.}\ \bibnamefont
  {Kadochkin}}, \bibinfo {author} {\bibfnamefont {V.~V.}\ \bibnamefont
  {Svetukhin}},\ and\ \bibinfo {author} {\bibfnamefont {A.~A.}\ \bibnamefont
  {Fotiadi}},\ }\href {https://doi.org/10.1103/PhysRevA.97.053828} {\bibfield
  {journal} {\bibinfo  {journal} {Physical Review A}\ }\textbf {\bibinfo
  {volume} {97}},\ \bibinfo {pages} {1} (\bibinfo {year} {2018})}\BibitemShut
  {NoStop}%
\bibitem [{\citenamefont {Dong}\ \emph {et~al.}(2021)\citenamefont {Dong},
  \citenamefont {Xiong}, \citenamefont {Phinney}, \citenamefont {Sun},
  \citenamefont {Jing}, \citenamefont {McLeod}, \citenamefont {Zhang},
  \citenamefont {Liu}, \citenamefont {Ruta}, \citenamefont {Gao}, \citenamefont
  {Dong}, \citenamefont {Pan}, \citenamefont {Edgar}, \citenamefont
  {Jarillo-Herrero}, \citenamefont {Levitov}, \citenamefont {Millis},
  \citenamefont {Fogler}, \citenamefont {Bandurin},\ and\ \citenamefont
  {Basov}}]{Dong2021}%
  \BibitemOpen
  \bibfield  {author} {\bibinfo {author} {\bibfnamefont {Y.}~\bibnamefont
  {Dong}}, \bibinfo {author} {\bibfnamefont {L.}~\bibnamefont {Xiong}},
  \bibinfo {author} {\bibfnamefont {I.~Y.}\ \bibnamefont {Phinney}}, \bibinfo
  {author} {\bibfnamefont {Z.}~\bibnamefont {Sun}}, \bibinfo {author}
  {\bibfnamefont {R.}~\bibnamefont {Jing}}, \bibinfo {author} {\bibfnamefont
  {A.~S.}\ \bibnamefont {McLeod}}, \bibinfo {author} {\bibfnamefont
  {S.}~\bibnamefont {Zhang}}, \bibinfo {author} {\bibfnamefont
  {S.}~\bibnamefont {Liu}}, \bibinfo {author} {\bibfnamefont {F.~L.}\
  \bibnamefont {Ruta}}, \bibinfo {author} {\bibfnamefont {H.}~\bibnamefont
  {Gao}}, \bibinfo {author} {\bibfnamefont {Z.}~\bibnamefont {Dong}}, \bibinfo
  {author} {\bibfnamefont {R.}~\bibnamefont {Pan}}, \bibinfo {author}
  {\bibfnamefont {J.~H.}\ \bibnamefont {Edgar}}, \bibinfo {author}
  {\bibfnamefont {P.}~\bibnamefont {Jarillo-Herrero}}, \bibinfo {author}
  {\bibfnamefont {L.~S.}\ \bibnamefont {Levitov}}, \bibinfo {author}
  {\bibfnamefont {A.~J.}\ \bibnamefont {Millis}}, \bibinfo {author}
  {\bibfnamefont {M.~M.}\ \bibnamefont {Fogler}}, \bibinfo {author}
  {\bibfnamefont {D.~A.}\ \bibnamefont {Bandurin}},\ and\ \bibinfo {author}
  {\bibfnamefont {D.~N.}\ \bibnamefont {Basov}},\ }\href
  {https://doi.org/10.1038/s41586-021-03640-x} {\bibfield  {journal} {\bibinfo
  {journal} {Nature}\ }\textbf {\bibinfo {volume} {594}},\ \bibinfo {pages}
  {513} (\bibinfo {year} {2021})},\ \Eprint {https://arxiv.org/abs/2103.10831}
  {2103.10831} \BibitemShut {NoStop}%
\bibitem [{\citenamefont {Cosme}\ \emph {et~al.}(2023)\citenamefont {Cosme},
  \citenamefont {Santos}, \citenamefont {Bizarro},\ and\ \citenamefont
  {Figueiredo}}]{Cosme2023TETHYS:Models}%
  \BibitemOpen
  \bibfield  {author} {\bibinfo {author} {\bibfnamefont {P.}~\bibnamefont
  {Cosme}}, \bibinfo {author} {\bibfnamefont {J.~S.}\ \bibnamefont {Santos}},
  \bibinfo {author} {\bibfnamefont {J.~P.}\ \bibnamefont {Bizarro}},\ and\
  \bibinfo {author} {\bibfnamefont {I.}~\bibnamefont {Figueiredo}},\ }\href
  {https://doi.org/10.1016/j.cpc.2022.108550} {\bibfield  {journal} {\bibinfo
  {journal} {Computer Physics Communications}\ }\textbf {\bibinfo {volume}
  {282}},\ \bibinfo {pages} {108550} (\bibinfo {year} {2023})}\BibitemShut
  {NoStop}%
\bibitem [{\citenamefont {Ooi}\ and\ \citenamefont {Tan}(2017)}]{Ooi2017}%
  \BibitemOpen
  \bibfield  {author} {\bibinfo {author} {\bibfnamefont {K.~J.~A.}\
  \bibnamefont {Ooi}}\ and\ \bibinfo {author} {\bibfnamefont {D.~T.~H.}\
  \bibnamefont {Tan}},\ }\href {https://doi.org/10.1098/rspa.2017.0433}
  {\bibfield  {journal} {\bibinfo  {journal} {Proceedings of the Royal Society
  A: Mathematical, Physical and Engineering Sciences}\ }\textbf {\bibinfo
  {volume} {473}},\ \bibinfo {pages} {20170433} (\bibinfo {year}
  {2017})}\BibitemShut {NoStop}%
\bibitem [{\citenamefont {Cox}\ and\ \citenamefont {{Garc{\'{i}}a de
  Abajo}}(2019)}]{Cox2019}%
  \BibitemOpen
  \bibfield  {author} {\bibinfo {author} {\bibfnamefont {J.~D.}\ \bibnamefont
  {Cox}}\ and\ \bibinfo {author} {\bibfnamefont {F.~J.}\ \bibnamefont
  {{Garc{\'{i}}a de Abajo}}},\ }\href
  {https://doi.org/10.1021/acs.accounts.9b00308} {\bibfield  {journal}
  {\bibinfo  {journal} {Accounts of Chemical Research}\ }\textbf {\bibinfo
  {volume} {52}},\ \bibinfo {pages} {2536} (\bibinfo {year}
  {2019})}\BibitemShut {NoStop}%
\bibitem [{\citenamefont {Han}\ \emph {et~al.}(2022)\citenamefont {Han},
  \citenamefont {Chin}, \citenamefont {Matschy}, \citenamefont {Poojali},
  \citenamefont {Seidl}, \citenamefont {Winnerl}, \citenamefont {Hafez},
  \citenamefont {Turchinovich}, \citenamefont {Kumar}, \citenamefont
  {Myers-Ward}, \citenamefont {Dejarld}, \citenamefont {Daniels}, \citenamefont
  {Drew}, \citenamefont {Murphy},\ and\ \citenamefont {Mittendorff}}]{Han2022}%
  \BibitemOpen
  \bibfield  {author} {\bibinfo {author} {\bibfnamefont {J.~W.}\ \bibnamefont
  {Han}}, \bibinfo {author} {\bibfnamefont {M.~L.}\ \bibnamefont {Chin}},
  \bibinfo {author} {\bibfnamefont {S.}~\bibnamefont {Matschy}}, \bibinfo
  {author} {\bibfnamefont {J.}~\bibnamefont {Poojali}}, \bibinfo {author}
  {\bibfnamefont {A.}~\bibnamefont {Seidl}}, \bibinfo {author} {\bibfnamefont
  {S.}~\bibnamefont {Winnerl}}, \bibinfo {author} {\bibfnamefont {H.~A.}\
  \bibnamefont {Hafez}}, \bibinfo {author} {\bibfnamefont {D.}~\bibnamefont
  {Turchinovich}}, \bibinfo {author} {\bibfnamefont {G.}~\bibnamefont {Kumar}},
  \bibinfo {author} {\bibfnamefont {R.~L.}\ \bibnamefont {Myers-Ward}},
  \bibinfo {author} {\bibfnamefont {M.~T.}\ \bibnamefont {Dejarld}}, \bibinfo
  {author} {\bibfnamefont {K.~M.}\ \bibnamefont {Daniels}}, \bibinfo {author}
  {\bibfnamefont {H.~D.}\ \bibnamefont {Drew}}, \bibinfo {author}
  {\bibfnamefont {T.~E.}\ \bibnamefont {Murphy}},\ and\ \bibinfo {author}
  {\bibfnamefont {M.}~\bibnamefont {Mittendorff}},\ }\href
  {https://doi.org/10.1002/adpr.202100218} {\bibfield  {journal} {\bibinfo
  {journal} {Advanced Photonics Research}\ }\textbf {\bibinfo {volume} {3}},\
  \bibinfo {pages} {2100218} (\bibinfo {year} {2022})}\BibitemShut {NoStop}%
\bibitem [{\citenamefont {Figueiredo}\ \emph {et~al.}(2022)\citenamefont
  {Figueiredo}, \citenamefont {Bizarro},\ and\ \citenamefont
  {Ter{\c{c}}as}}]{Figueiredo2020}%
  \BibitemOpen
  \bibfield  {author} {\bibinfo {author} {\bibfnamefont {J.~L.}\ \bibnamefont
  {Figueiredo}}, \bibinfo {author} {\bibfnamefont {J.~P.~S.}\ \bibnamefont
  {Bizarro}},\ and\ \bibinfo {author} {\bibfnamefont {H.}~\bibnamefont
  {Ter{\c{c}}as}},\ }\href {https://doi.org/10.1088/1367-2630/ac5132}
  {\bibfield  {journal} {\bibinfo  {journal} {New Journal of Physics}\ }\textbf
  {\bibinfo {volume} {24}},\ \bibinfo {pages} {023026} (\bibinfo {year}
  {2022})},\ \Eprint {https://arxiv.org/abs/2012.15148} {2012.15148}
  \BibitemShut {NoStop}%
\bibitem [{\citenamefont {Chaves}\ \emph {et~al.}(2017)\citenamefont {Chaves},
  \citenamefont {Peres}, \citenamefont {Smirnov}, \citenamefont
  {Asger~Mortensen},\ and\ \citenamefont {Mortensen}}]{Chaves2017}%
  \BibitemOpen
  \bibfield  {author} {\bibinfo {author} {\bibfnamefont {A.~J.}\ \bibnamefont
  {Chaves}}, \bibinfo {author} {\bibfnamefont {N.~M.~R.}\ \bibnamefont
  {Peres}}, \bibinfo {author} {\bibfnamefont {G.}~\bibnamefont {Smirnov}},
  \bibinfo {author} {\bibfnamefont {N.}~\bibnamefont {Asger~Mortensen}},\ and\
  \bibinfo {author} {\bibfnamefont {N.~A.}\ \bibnamefont {Mortensen}},\ }\href
  {https://doi.org/10.1103/PhysRevB.96.195438} {\bibfield  {journal} {\bibinfo
  {journal} {Physical Review B}\ }\textbf {\bibinfo {volume} {96}},\ \bibinfo
  {pages} {195438} (\bibinfo {year} {2017})}\BibitemShut {NoStop}%
\bibitem [{\citenamefont {Avron}(1998)}]{Avron1998OddViscosity}%
  \BibitemOpen
  \bibfield  {author} {\bibinfo {author} {\bibfnamefont {J.~E.}\ \bibnamefont
  {Avron}},\ }\href {https://doi.org/10.1023/a:1023084404080} {\bibfield
  {journal} {\bibinfo  {journal} {Journal of Statistical Physics}\ }\textbf
  {\bibinfo {volume} {92}},\ \bibinfo {pages} {543} (\bibinfo {year}
  {1998})}\BibitemShut {NoStop}%
\bibitem [{\citenamefont {Narozhny}\ and\ \citenamefont
  {Sch{\"{u}}tt}(2019)}]{Narozhny2019MagnetohydrodynamicsViscosities}%
  \BibitemOpen
  \bibfield  {author} {\bibinfo {author} {\bibfnamefont {B.~N.}\ \bibnamefont
  {Narozhny}}\ and\ \bibinfo {author} {\bibfnamefont {M.}~\bibnamefont
  {Sch{\"{u}}tt}},\ }\href {https://doi.org/10.1103/PhysRevB.100.035125}
  {\bibfield  {journal} {\bibinfo  {journal} {Physical Review B}\ }\textbf
  {\bibinfo {volume} {100}},\ \bibinfo {pages} {035125} (\bibinfo {year}
  {2019})}\BibitemShut {NoStop}%
\bibitem [{\citenamefont {Chen}\ and\ \citenamefont
  {Zhu}(2022)}]{Chen2022ViscosityElectrons}%
  \BibitemOpen
  \bibfield  {author} {\bibinfo {author} {\bibfnamefont {W.}~\bibnamefont
  {Chen}}\ and\ \bibinfo {author} {\bibfnamefont {W.}~\bibnamefont {Zhu}},\
  }\href {https://doi.org/10.1103/PhysRevB.106.014205} {\bibfield  {journal}
  {\bibinfo  {journal} {Physical Review B}\ }\textbf {\bibinfo {volume}
  {106}},\ \bibinfo {pages} {014205} (\bibinfo {year} {2022})}\BibitemShut
  {NoStop}%
\bibitem [{\citenamefont {Friedman}\ \emph {et~al.}(2022)\citenamefont
  {Friedman}, \citenamefont {Cook},\ and\ \citenamefont
  {Lucas}}]{Friedman2022HydrodynamicsGroup}%
  \BibitemOpen
  \bibfield  {author} {\bibinfo {author} {\bibfnamefont {A.~J.}\ \bibnamefont
  {Friedman}}, \bibinfo {author} {\bibfnamefont {C.~Q.}\ \bibnamefont {Cook}},\
  and\ \bibinfo {author} {\bibfnamefont {A.}~\bibnamefont {Lucas}},\ }\href
  {http://arxiv.org/abs/2202.08269} {\bibfield  {journal} {\bibinfo  {journal}
  {arXiv Strongly Correlated Electrons}\ ,\ \bibinfo {pages} {1}} (\bibinfo
  {year} {2022})}\BibitemShut {NoStop}%
\bibitem [{\citenamefont {Manfredi}\ and\ \citenamefont
  {Haas}(2001)}]{Manfredi2001Self-consistentGas}%
  \BibitemOpen
  \bibfield  {author} {\bibinfo {author} {\bibfnamefont {G.}~\bibnamefont
  {Manfredi}}\ and\ \bibinfo {author} {\bibfnamefont {F.}~\bibnamefont
  {Haas}},\ }\href {https://doi.org/10.1103/PhysRevB.64.075316} {\bibfield
  {journal} {\bibinfo  {journal} {Physical Review B}\ }\textbf {\bibinfo
  {volume} {64}},\ \bibinfo {pages} {075316} (\bibinfo {year}
  {2001})}\BibitemShut {NoStop}%
\bibitem [{\citenamefont {Haas}(2011)}]{Haas2011}%
  \BibitemOpen
  \bibfield  {author} {\bibinfo {author} {\bibfnamefont {F.}~\bibnamefont
  {Haas}},\ }\href {https://doi.org/10.1007/978-1-4419-8201-8} {\emph {\bibinfo
  {title} {{Quantum Plasmas}}}},\ \bibinfo {series} {Springer Series on Atomic,
  Optical, and Plasma Physics}, Vol.~\bibinfo {volume} {65}\ (\bibinfo
  {publisher} {Springer New York},\ \bibinfo {address} {New York, NY},\
  \bibinfo {year} {2011})\BibitemShut {NoStop}%
\bibitem [{\citenamefont {Shur}(1990)}]{Shur1990}%
  \BibitemOpen
  \bibfield  {author} {\bibinfo {author} {\bibfnamefont {M.}~\bibnamefont
  {Shur}},\ }\href@noop {} {\emph {\bibinfo {title} {{Physics of Semiconductor
  Devices}}}}\ (\bibinfo  {publisher} {Prentice Hall},\ \bibinfo {year}
  {1990})\BibitemShut {NoStop}%
\bibitem [{\citenamefont {Svintsov}\ \emph {et~al.}(2013)\citenamefont
  {Svintsov}, \citenamefont {Vyurkov}, \citenamefont {Ryzhii},\ and\
  \citenamefont {Otsuji}}]{Svintsov2013}%
  \BibitemOpen
  \bibfield  {author} {\bibinfo {author} {\bibfnamefont {D.}~\bibnamefont
  {Svintsov}}, \bibinfo {author} {\bibfnamefont {V.}~\bibnamefont {Vyurkov}},
  \bibinfo {author} {\bibfnamefont {V.}~\bibnamefont {Ryzhii}},\ and\ \bibinfo
  {author} {\bibfnamefont {T.}~\bibnamefont {Otsuji}},\ }\href
  {https://doi.org/10.1103/PhysRevB.88.245444} {\bibfield  {journal} {\bibinfo
  {journal} {Physical Review B - Condensed Matter and Materials Physics}\
  }\textbf {\bibinfo {volume} {88}},\ \bibinfo {pages} {245444} (\bibinfo
  {year} {2013})}\BibitemShut {NoStop}%
\bibitem [{\citenamefont {Ali}\ \emph {et~al.}(2007)\citenamefont {Ali},
  \citenamefont {Moslem}, \citenamefont {Shukla},\ and\ \citenamefont
  {Kourakis}}]{Ali2007}%
  \BibitemOpen
  \bibfield  {author} {\bibinfo {author} {\bibfnamefont {S.}~\bibnamefont
  {Ali}}, \bibinfo {author} {\bibfnamefont {W.~M.}\ \bibnamefont {Moslem}},
  \bibinfo {author} {\bibfnamefont {P.~K.}\ \bibnamefont {Shukla}},\ and\
  \bibinfo {author} {\bibfnamefont {I.}~\bibnamefont {Kourakis}},\ }\href
  {https://doi.org/10.1016/j.physleta.2007.05.073} {\bibfield  {journal}
  {\bibinfo  {journal} {Physics Letters, Section A: General, Atomic and Solid
  State Physics}\ }\textbf {\bibinfo {volume} {366}},\ \bibinfo {pages} {606}
  (\bibinfo {year} {2007})}\BibitemShut {NoStop}%
\bibitem [{\citenamefont {Haas}\ and\ \citenamefont
  {Kourakis}(2015)}]{Haas2015}%
  \BibitemOpen
  \bibfield  {author} {\bibinfo {author} {\bibfnamefont {F.}~\bibnamefont
  {Haas}}\ and\ \bibinfo {author} {\bibfnamefont {I.}~\bibnamefont
  {Kourakis}},\ }\bibfield  {journal} {\bibinfo  {journal} {Plasma Physics and
  Controlled Fusion}\ }\textbf {\bibinfo {volume} {57}},\ \href
  {https://doi.org/10.1088/0741-3335/57/4/044006}
  {10.1088/0741-3335/57/4/044006} (\bibinfo {year} {2015})\BibitemShut
  {NoStop}%
\bibitem [{\citenamefont {Sagdeev}\ and\ \citenamefont
  {Galeev}(1969)}]{Sagdeev1969NonlinearTheory}%
  \BibitemOpen
  \bibfield  {author} {\bibinfo {author} {\bibfnamefont {R.~Z.}\ \bibnamefont
  {Sagdeev}}\ and\ \bibinfo {author} {\bibfnamefont {A.~A.}\ \bibnamefont
  {Galeev}},\ }\href@noop {} {\emph {\bibinfo {title} {{Nonlinear Plasma
  Theory}}}}\ (\bibinfo  {publisher} {W. A. Benjamin, Inc.},\ \bibinfo
  {address} {New York, NY},\ \bibinfo {year} {1969})\BibitemShut {NoStop}%
\bibitem [{\citenamefont {Sagdeev}\ \emph {et~al.}(1988)\citenamefont
  {Sagdeev}, \citenamefont {Usikov},\ and\ \citenamefont
  {Zaslavsky}}]{Sagdeev1988NonlinearChaos}%
  \BibitemOpen
  \bibfield  {author} {\bibinfo {author} {\bibfnamefont {R.~Z.}\ \bibnamefont
  {Sagdeev}}, \bibinfo {author} {\bibfnamefont {D.~A.}\ \bibnamefont
  {Usikov}},\ and\ \bibinfo {author} {\bibfnamefont {G.~M.}\ \bibnamefont
  {Zaslavsky}},\ }\href@noop {} {\emph {\bibinfo {title} {{Nonlinear physics:
  from the pendulum to turbulence and chaos}}}}\ (\bibinfo  {publisher}
  {Harwood Academic Publishers},\ \bibinfo {year} {1988})\BibitemShut {NoStop}%
\bibitem [{\citenamefont {Taniuti}\ and\ \citenamefont
  {Wei}(1968)}]{Taniuti1968ReductiveI}%
  \BibitemOpen
  \bibfield  {author} {\bibinfo {author} {\bibfnamefont {T.}~\bibnamefont
  {Taniuti}}\ and\ \bibinfo {author} {\bibfnamefont {C.~C.}\ \bibnamefont
  {Wei}},\ }\href {https://doi.org/10.1143/JPSJ.24.941} {\bibfield  {journal}
  {\bibinfo  {journal} {Journal of the Physical Society of Japan}\ }\textbf
  {\bibinfo {volume} {24}},\ \bibinfo {pages} {941} (\bibinfo {year}
  {1968})}\BibitemShut {NoStop}%
\bibitem [{\citenamefont {Taniuti}\ and\ \citenamefont
  {Yajima}(1969)}]{Taniuti1969PerturbationI}%
  \BibitemOpen
  \bibfield  {author} {\bibinfo {author} {\bibfnamefont {T.}~\bibnamefont
  {Taniuti}}\ and\ \bibinfo {author} {\bibfnamefont {N.}~\bibnamefont
  {Yajima}},\ }\href {https://doi.org/10.1063/1.1664975} {\bibfield  {journal}
  {\bibinfo  {journal} {Journal of Mathematical Physics}\ }\textbf {\bibinfo
  {volume} {10}},\ \bibinfo {pages} {1369} (\bibinfo {year}
  {1969})}\BibitemShut {NoStop}%
\bibitem [{\citenamefont {Taniuti}\ and\ \citenamefont
  {Nishihara}(1983)}]{Taniuti1983NonlinearWaves}%
  \BibitemOpen
  \bibfield  {author} {\bibinfo {author} {\bibfnamefont {T.}~\bibnamefont
  {Taniuti}}\ and\ \bibinfo {author} {\bibfnamefont {K.}~\bibnamefont
  {Nishihara}},\ }\href@noop {} {\emph {\bibinfo {title} {{Nonlinear waves}}}}\
  (\bibinfo  {publisher} {Pitman Advanced Publishing Program},\ \bibinfo {year}
  {1983})\BibitemShut {NoStop}%
\bibitem [{\citenamefont {Leblond}(2008)}]{Leblond2008TheApplications}%
  \BibitemOpen
  \bibfield  {author} {\bibinfo {author} {\bibfnamefont {H.}~\bibnamefont
  {Leblond}},\ }\bibfield  {journal} {\bibinfo  {journal} {Journal of Physics
  B: Atomic, Molecular and Optical Physics}\ }\textbf {\bibinfo {volume}
  {41}},\ \href {https://doi.org/10.1088/0953-4075/41/4/043001}
  {10.1088/0953-4075/41/4/043001} (\bibinfo {year} {2008})\BibitemShut
  {NoStop}%
\bibitem [{\citenamefont {Bartuccelli}\ \emph {et~al.}(1985)\citenamefont
  {Bartuccelli}, \citenamefont {Carbonaro},\ and\ \citenamefont
  {Muto}}]{Bartuccelli1985Kadomtsev-Petviashvili-BurgersWaves}%
  \BibitemOpen
  \bibfield  {author} {\bibinfo {author} {\bibfnamefont {M.}~\bibnamefont
  {Bartuccelli}}, \bibinfo {author} {\bibfnamefont {P.}~\bibnamefont
  {Carbonaro}},\ and\ \bibinfo {author} {\bibfnamefont {V.}~\bibnamefont
  {Muto}},\ }\href {https://doi.org/10.1007/BF02722453} {\bibfield  {journal}
  {\bibinfo  {journal} {Lettere Al Nuovo Cimento Series 2}\ }\textbf {\bibinfo
  {volume} {42}},\ \bibinfo {pages} {279} (\bibinfo {year} {1985})}\BibitemShut
  {NoStop}%
\bibitem [{\citenamefont {Ghosh}\ and\ \citenamefont
  {Sahu}(2019)}]{Ghosh2019NonlinearPlasma}%
  \BibitemOpen
  \bibfield  {author} {\bibinfo {author} {\bibfnamefont {N.}~\bibnamefont
  {Ghosh}}\ and\ \bibinfo {author} {\bibfnamefont {B.}~\bibnamefont {Sahu}},\
  }\href {https://doi.org/10.1088/0253-6102/71/2/237} {\bibfield  {journal}
  {\bibinfo  {journal} {Communications in Theoretical Physics}\ }\textbf
  {\bibinfo {volume} {71}},\ \bibinfo {pages} {237} (\bibinfo {year}
  {2019})}\BibitemShut {NoStop}%
\bibitem [{\citenamefont {Misra}\ and\ \citenamefont
  {Sahu}(2015)}]{Misra2015MultidimensionalPlasmas}%
  \BibitemOpen
  \bibfield  {author} {\bibinfo {author} {\bibfnamefont {A.~P.}\ \bibnamefont
  {Misra}}\ and\ \bibinfo {author} {\bibfnamefont {B.}~\bibnamefont {Sahu}},\
  }\href {https://doi.org/10.1016/j.physa.2014.11.045} {\bibfield  {journal}
  {\bibinfo  {journal} {Physica A: Statistical Mechanics and its Applications}\
  }\textbf {\bibinfo {volume} {421}},\ \bibinfo {pages} {269} (\bibinfo {year}
  {2015})}\BibitemShut {NoStop}%
\bibitem [{\citenamefont {Seadawy}(2017)}]{Seadawy2017IonPlasma}%
  \BibitemOpen
  \bibfield  {author} {\bibinfo {author} {\bibfnamefont {A.~R.}\ \bibnamefont
  {Seadawy}},\ }\href {https://doi.org/10.1002/mma.4081} {\bibfield  {journal}
  {\bibinfo  {journal} {Mathematical Methods in the Applied Sciences}\ }\textbf
  {\bibinfo {volume} {40}},\ \bibinfo {pages} {1598} (\bibinfo {year}
  {2017})}\BibitemShut {NoStop}%
\bibitem [{\citenamefont {Canosa}\ and\ \citenamefont
  {Gazdag}(1977)}]{Canosa1977TheEquation}%
  \BibitemOpen
  \bibfield  {author} {\bibinfo {author} {\bibfnamefont {J.}~\bibnamefont
  {Canosa}}\ and\ \bibinfo {author} {\bibfnamefont {J.}~\bibnamefont
  {Gazdag}},\ }\href {https://doi.org/10.1016/0021-9991(77)90070-5} {\bibfield
  {journal} {\bibinfo  {journal} {Journal of Computational Physics}\ }\textbf
  {\bibinfo {volume} {23}},\ \bibinfo {pages} {393} (\bibinfo {year}
  {1977})}\BibitemShut {NoStop}%
\bibitem [{\citenamefont {Bona}\ and\ \citenamefont
  {Schonbek}(1985)}]{Bona1985Travelling-waveEquation}%
  \BibitemOpen
  \bibfield  {author} {\bibinfo {author} {\bibfnamefont {J.~L.}\ \bibnamefont
  {Bona}}\ and\ \bibinfo {author} {\bibfnamefont {M.~E.}\ \bibnamefont
  {Schonbek}},\ }\href {https://doi.org/10.1017/S0308210500020783} {\bibfield
  {journal} {\bibinfo  {journal} {Proceedings of the Royal Society of
  Edinburgh: Section A Mathematics}\ }\textbf {\bibinfo {volume} {101}},\
  \bibinfo {pages} {207} (\bibinfo {year} {1985})}\BibitemShut {NoStop}%
\bibitem [{\citenamefont {Jeffrey}\ and\ \citenamefont
  {Xu}(1989)}]{Jeffrey1989ExactEquation}%
  \BibitemOpen
  \bibfield  {author} {\bibinfo {author} {\bibfnamefont {A.}~\bibnamefont
  {Jeffrey}}\ and\ \bibinfo {author} {\bibfnamefont {S.}~\bibnamefont {Xu}},\
  }\href {https://doi.org/10.1016/0165-2125(89)90026-7} {\bibfield  {journal}
  {\bibinfo  {journal} {Wave Motion}\ }\textbf {\bibinfo {volume} {11}},\
  \bibinfo {pages} {559} (\bibinfo {year} {1989})}\BibitemShut {NoStop}%
\bibitem [{\citenamefont {Feng}(2003)}]{Feng2003ExactEquation}%
  \BibitemOpen
  \bibfield  {author} {\bibinfo {author} {\bibfnamefont {Z.}~\bibnamefont
  {Feng}},\ }\href {https://doi.org/10.1016/S0165-2125(03)00023-4} {\bibfield
  {journal} {\bibinfo  {journal} {Wave Motion}\ }\textbf {\bibinfo {volume}
  {38}},\ \bibinfo {pages} {109} (\bibinfo {year} {2003})}\BibitemShut
  {NoStop}%
\bibitem [{\citenamefont {Ablowitz}\ and\ \citenamefont
  {Zeppetella}(1979)}]{ABLOWITZ1979}%
  \BibitemOpen
  \bibfield  {author} {\bibinfo {author} {\bibfnamefont {M.}~\bibnamefont
  {Ablowitz}}\ and\ \bibinfo {author} {\bibfnamefont {A.}~\bibnamefont
  {Zeppetella}},\ }\href {https://doi.org/10.1016/S0092-8240(79)80020-8}
  {\bibfield  {journal} {\bibinfo  {journal} {Bulletin of Mathematical
  Biology}\ }\textbf {\bibinfo {volume} {41}},\ \bibinfo {pages} {835}
  (\bibinfo {year} {1979})}\BibitemShut {NoStop}%
\bibitem [{\citenamefont {Kourakis}\ \emph {et~al.}(2012)\citenamefont
  {Kourakis}, \citenamefont {Sultana},\ and\ \citenamefont
  {Verheest}}]{Kourakis2012}%
  \BibitemOpen
  \bibfield  {author} {\bibinfo {author} {\bibfnamefont {I.}~\bibnamefont
  {Kourakis}}, \bibinfo {author} {\bibfnamefont {S.}~\bibnamefont {Sultana}},\
  and\ \bibinfo {author} {\bibfnamefont {F.}~\bibnamefont {Verheest}},\ }\href
  {https://doi.org/10.1007/s10509-011-0958-5} {\bibfield  {journal} {\bibinfo
  {journal} {Astrophysics and Space Science}\ }\textbf {\bibinfo {volume}
  {338}},\ \bibinfo {pages} {245} (\bibinfo {year} {2012})}\BibitemShut
  {NoStop}%
\bibitem [{\citenamefont {Zakharov}\ and\ \citenamefont
  {Ostrovsky}(2009)}]{Zakharov2009ModulationBeginning}%
  \BibitemOpen
  \bibfield  {author} {\bibinfo {author} {\bibfnamefont {V.~E.}\ \bibnamefont
  {Zakharov}}\ and\ \bibinfo {author} {\bibfnamefont {L.~A.}\ \bibnamefont
  {Ostrovsky}},\ }\href {https://doi.org/10.1016/j.physd.2008.12.002}
  {\bibfield  {journal} {\bibinfo  {journal} {Physica D: Nonlinear Phenomena}\
  }\textbf {\bibinfo {volume} {238}},\ \bibinfo {pages} {540} (\bibinfo {year}
  {2009})}\BibitemShut {NoStop}%
\bibitem [{\citenamefont {Kourakis}\ \emph {et~al.}(2007)\citenamefont
  {Kourakis}, \citenamefont {Lazarides},\ and\ \citenamefont
  {Tsironis}}]{Kourakis2007}%
  \BibitemOpen
  \bibfield  {author} {\bibinfo {author} {\bibfnamefont {I.}~\bibnamefont
  {Kourakis}}, \bibinfo {author} {\bibfnamefont {N.}~\bibnamefont
  {Lazarides}},\ and\ \bibinfo {author} {\bibfnamefont {G.~P.}\ \bibnamefont
  {Tsironis}},\ }\href {https://doi.org/10.1103/PhysRevE.75.067601} {\bibfield
  {journal} {\bibinfo  {journal} {Physical Review E - Statistical, Nonlinear,
  and Soft Matter Physics}\ }\textbf {\bibinfo {volume} {75}},\ \bibinfo
  {pages} {2} (\bibinfo {year} {2007})},\ \Eprint
  {https://arxiv.org/abs/0612615} {0612615} \BibitemShut {NoStop}%
\bibitem [{\citenamefont {Peregrine}(1983)}]{Peregrine1983WaterSolutions}%
  \BibitemOpen
  \bibfield  {author} {\bibinfo {author} {\bibfnamefont {D.~H.}\ \bibnamefont
  {Peregrine}},\ }\href {https://doi.org/10.1017/S0334270000003891} {\bibfield
  {journal} {\bibinfo  {journal} {The Journal of the Australian Mathematical
  Society. Series B. Applied Mathematics}\ }\textbf {\bibinfo {volume} {25}},\
  \bibinfo {pages} {16} (\bibinfo {year} {1983})}\BibitemShut {NoStop}%
\bibitem [{\citenamefont {Zakharov}\ and\ \citenamefont
  {Gelash}(2013)}]{Zakharov2013NonlinearInstability}%
  \BibitemOpen
  \bibfield  {author} {\bibinfo {author} {\bibfnamefont {V.~E.}\ \bibnamefont
  {Zakharov}}\ and\ \bibinfo {author} {\bibfnamefont {A.~A.}\ \bibnamefont
  {Gelash}},\ }\href {https://doi.org/10.1103/PhysRevLett.111.054101}
  {\bibfield  {journal} {\bibinfo  {journal} {Physical Review Letters}\
  }\textbf {\bibinfo {volume} {111}},\ \bibinfo {pages} {1} (\bibinfo {year}
  {2013})}\BibitemShut {NoStop}%
\bibitem [{\citenamefont {Zabusky}\ and\ \citenamefont
  {Kruskal}(1965)}]{zabusky1965}%
  \BibitemOpen
  \bibfield  {author} {\bibinfo {author} {\bibfnamefont {N.~J.}\ \bibnamefont
  {Zabusky}}\ and\ \bibinfo {author} {\bibfnamefont {M.~D.}\ \bibnamefont
  {Kruskal}},\ }\href {https://doi.org/10.1103/PhysRevLett.15.240} {\bibfield
  {journal} {\bibinfo  {journal} {Phys. Rev. Lett.}\ }\textbf {\bibinfo
  {volume} {15}},\ \bibinfo {pages} {240} (\bibinfo {year} {1965})}\BibitemShut
  {NoStop}%
\end{thebibliography}%

\end{document}